\documentclass{aa}  
\usepackage[dvipsnames]{xcolor} 
\usepackage{graphicx}
\usepackage{caption}
\usepackage{txfonts}
\usepackage{amsmath}
\usepackage{mhchem}
\usepackage{natbib}
\usepackage{lipsum}
\usepackage{subcaption} 
\usepackage{lscape} 
\usepackage{placeins}
\usepackage[colorlinks=true,linkcolor=blue, citecolor=blue]{hyperref}%
\begin{document}

   \title{Hunting for methanol in the water rich, planet forming disk around HL Tau}

   \author{A. Soave\inst{1,2}
          \and
          M. Leemker\inst{3}
          \and
          S. Facchini\inst{3}
          \and
          L. Maud\inst{4}
          \and
          K. L. J. Rygl\inst{5}
          \and
          L. Testi\inst{1,6}
          }

   \institute{Alma Mater Studiorum – Università di Bologna, Dipartimento di Fisica e Astronomia “Augusto Righi”, Via Gobetti 93/2, I-40129, Bologna, Italy\\
              \email{alessandro.soave2@unibo.it}
        \and
            INAF – Osservatorio di Astrofisica e Scienza dello Spazio, Via P. Gobetti 93/3, 40129 Bologna, Italy 
        \and 
            Dipartimento di Fisica, Università degli Studi di Milano, Via Celoria 16, 20133 Milano, Italy 
        \and 
            European Southern Observatory, Karl-Schwarzschild Str. 2, 85748 Garching bei München, Germany
        \and
            INAF-Istituto di Radioastronomia \& Italian ALMA Regional Centre, Via P. Gobetti 101, I-40129 Bologna, Italy
        \and
            INAF-Osservatorio Astrofisico di Arcetri, Largo E. Fermi 5, I-50125 Firenze, Italy}
   \date{Received 9 December 2025; accepted 20 February 2026}

 
  \abstract
    {Methanol, the simplest complex organic molecule found in space, is considered a key compound necessary for the formation of chemical species of prebiotic interest. Methanol detections in protoplanetary disks remain scarce, even though it is frequently detected in the material surrounding other Young Stellar Objects.}
   {We investigate the presence of methanol in the protoplanetary disk around the HL Tau protostar, motivated by the detection of spatially resolved warm water emission.}
   {Given the similar volatility of methanol and water, thermally desorbed gas-phase methanol is expected to emit from the same region of the HL Tau disk where water vapor has been observed. Accordingly, we selected and imaged the most promising ALMA archival observations to search for rotational methanol lines.}
   {We found no methanol emission in the analyzed archival datasets. Assuming optically thin emission and LTE, we derive stringent upper limits on the methanol column density for different excitation temperatures: < $7.2 \times 10^{14} \ \mathrm{cm^{-2}}$ at 100 K and < $1.8 \times 10^{15} \ \mathrm{cm^{-2}}$ at 200 K, assuming a circular emitting region with a radius of 17 au ($\sim$ 0.12$''$). Furthermore, we obtain a stringent upper limit on the methanol-to-water column density ratio (< $0.55 \times 10^{-3}$ at 100 K and < $1.4 \times 10^{-3}$ at 200 K), which is, on average, an order of magnitude lower than the values measured for other Young Stellar Objects and Solar System comets.}
   {We argue that the most likely explanation for the methanol non-detection in HL Tau is the presence of optically thick dust in the central region of the disk, which obscures part of the methanol emission. The upper limit on the methanol-to-water ratio in the HL Tau disk is at least an order of magnitude smaller than most clouds, YSOs and comets, possibly due to radiative transfer and/or excitation effects, or due to a different chemical evolution compared to the other sources.}


   \maketitle
%

\section{Introduction}\label{sec:intro}

Formed inside the cold and dense environment of the Interstellar Medium, Complex Organic Molecules (COMs) are molecules containing both carbon and hydrogen with at least six atoms \citep{Herbst_2009, Ceccarelli_2023}. Among them, methanol (\ce{CH3OH}) -the simplest possible alcohol- is of great interest from a prebiotic point of view. Indeed, laboratory experiments have shown that photochemical reactions of UV-irradiated \ce{CH3OH} molecules can lead to the assembly of more complex organic compounds \citep{Oberg_2009, Fedoseev_2016, Chuang_2017}. Nowadays, emission lines tracing gas-phase methanol and absorption features tracing methanol ices, as well as other COMs, are commonly detected in a wide range of astrophysical sources, ranging from molecular clouds in star-forming regions to Class II Young Stellar Objects (YSO; \citealt{Boogert_2015, Chen_2023, Jorgensen_2020, Law_2025, McClure_2023, McGuire_2022, Pontoppidan_2023, Oberg_2023, Scibelli_2020}). Leveraging the new revolutionizing capabilities of the ALMA radiointerferometer, after the first detection of methyl cyanide in the MWC~480 disk (\ce{CH3CN}; \citealt{Oberg_2015}), COMs have started to be detected also in protoplanetary disks. Nevertheless, despite being ubiquitously present in younger YSOs, gas-phase \ce{CH_{3}OH} is still an elusive species, only found in seven protoplanetary disks to date \citep{Walsh_2016, vantHoff_2018, VanDerMarel_2021, Booth_2021, Booth_2023, Booth_2025a, Booth_2025b}. \par

This scarce detection rate of gas-phase \ce{CH_{3}OH} in protoplanetary disks can be attributed to the fact that the bulk of methanol is locked onto the icy surface of dust grains almost throughout the disk \citep{Walsh_2014} and thermally desorbes at the methanol snowline close to the protostar. The latter is the midplane location of the disk where temperatures reach the \ce{CH3OH} desorption temperature of $\sim$ 120 K \citep{Penteado_2017, Minissale_2022}. Recent studies have argued that, at least for warmer protoplanetary disks around Herbig Ae/Be stars, the \ce{CH3OH} reservoir is inherited from the natal molecular cloud \citep{Booth_2021}, where methanol is formed efficiently on the surface of dust grains via hydrogenation of carbon monoxide (CO; \citealt{Fuchs_2009}) and/or via radical-molecule H-atom abstraction reaction \citep{Santos_2022}. Moreover, recent evidence from a few disks suggests that the bulk of ices is inherited from the earliest phases of star and planet formation without undergoing extensive chemical processing. Indeed, observations of freshly sublimated water, whose deuteration level is a sensitive tracer of chemical processing in ices, in the L1551~IRS5 Class I source and in the V883~Ori disk reveal isotopic ratios similar to those found in protostellar envelopes  \citep{Andreu_2023, Tobin_2023, Leemker_2025_Nature}. Since there is no efficient methanol gas-phase formation pathway and the temperatures in the disk are too high to allow \ce{CO} to freeze out, it follows that the gas-phase \ce{CH3OH} must be originated from the sublimation of these pristine ices \citep{Garrod_2006, Geppert_2006}. \par

From an observational point of view, \ce{CH3OH} was detected for the first time in the TW~Hya protoplanetary disk by \cite{Walsh_2016}. Subsequent observations of methanol lines, supported by thermochemical modeling, indicate that non-thermal desorption processes, primarily photodesorption, are responsible for releasing \ce{CH3OH} into the gas phase \citep{Ilee+25}. After that, methanol emission has also been found in the V883 Ori protoplanetary disk, a YSO whose FU Orionis type-star is undergoing an accretion burst, shifting the water snowline out to $\sim$ 80 au in the outer disk \citep{vantHoff_2018, Lee_2019, Leemker_2021, Tobin_2023, Lee_2024, Jeong_2025, Wang_2025, Zeng_2025}. In this case, it is believed that the molecules detected in gas phase within the snowline are directly tracing the thermally sublimated ice coating the dust grains. Methanol has been found also in several transition protoplanetary disks around Herbig stars: HD~100546 \citep{Booth_2021, Booth_2024, Evans_2025}, IRS~48 (\citealt{VanDerMarel_2021, Brunken_2022, Booth_2024b, Temmink_2024}), HD~169142 \citep{Booth_2023}, HD~100453 \citep{Booth_2025a, Booth_2025b} and CQ~Tau \citep{Booth_2025b}. In these Herbig sources with \ce{CH3OH} detections, the desorption mechanism is thermal, motivated by the high temperatures reached by the directly irradiated cavity walls and, possibly, by the subsequent vertical cycling of the grains \citep{VanDerMarel_2021}. Stringent upper limits on the methanol column density have also been derived for the full disks around the Herbig stars HD 163296 and the MWC 480 \citep{Carney_2019, Yamato+24}. Similarly, tight constraints on the methanol budget in the younger DG~Tau~A disk, which is still in a transitional stage between a Class~I and II YSO, have been derived by \cite{Podio_2019}. \par

To better constrain the chemical processing and evolution of the methanol budget inside protoplanetary disks, it is often useful to compare its abundance to the one of other closely related molecules, such as water (\ce{H2O}). This is the case since methanol and water have a similar volatility, with an approximate desorption temperature of 120 and 150 K, respectively \citep{Penteado_2017, Minissale_2022}. It follows that methanol and water sublimate into gas-phase from the icy surface of dust grains in the same region of the disk, and that, consequently, their snowlines would be close to one another. Moreover, the methanol-to-water column density ratio is a useful diagnostic to study the degree of chemical processing after ice sublimation during the evolution of YSOs, from young still deeply embedded protostars to Class II disks. This ratio is indeed predicted to decrease over time due the combination of two effects: firstly,  \ce{CH3OH} is formed efficiently just on the surface of dust grains in ISM, while for water other gas formation routes may be effective \citep{van_Dishoeck_2013}, and secondly methanol is photo-dissociated two times faster than water \citep{Heays_2017}. \par 
As a consequence of the tight relation between methanol and water desorption temperature, protoplanetary disks where gas-phase water is detected could potentially show signs of methanol emission. As such, the HL Tau protoplanetary disk is a promising candidate to search for methanol, due to the recent discovery of three spectrally and spatially resolved \ce{H2O} lines, emitting primarily from within 17 au from the central star and likely tracing warm water \citep{Facchini_2024, Leemker_2025_A&A}. HL Tau is a young protostar ($\lesssim$ 1 Myr) located in the Taurus star forming region at a distance of $\sim$ 140 pc \citep{Galli_2018}. Its mass of 2.1 $M_{\odot}$ was estimated by \cite{Yen_2019} using dynamical measurements. The protoplanetary disk around HL Tau, which was the first ever to be imaged at high angular resolution with ALMA \citep{ALMA_long_baseline_2015}, is rich in substructures, showing a multitude of rings and gaps \citep{Stephens_2023}. The HL Tau system is classified as a YSO at a transition phase between Class I-II, due to the still ongoing interaction between the envelope and the protoplanetary disk, as traced also by the presence of a streamer and of an accretion shock seen with S-bearing molecules \citep{Yen_2019, Garufi_2022, Leemker_2025_A&A}. \par 
In this paper, we present our analysis of ALMA archival observations, in order to constrain the methanol budget inside the HL Tau protoplanetary disk. After introducing the methods and the ALMA observations in Sect.~\ref{sec:ALMA_archive}, Sect.~\ref{sec:result} shows the methanol non-detections and the derived upper limits on the methanol column density. Then, in Sect.~\ref{sec:discussion} we benchmark the HL Tau methanol-to-water column density ratio with the one measured for other YSOs and comets. We also compare the upper limit on the methanol abundance found in HL Tau with the one measured in disks with \ce{CH3OH} detections, taking advantage of available DALI thermochemical models. Additionally, we search for possible correlations between the sulfur monoxide (\ce{SO}) to methanol line flux ratio and the stellar luminosity in protoplanetary disks. This study allows us to connect the organic budget and the chemical complexity of the disk with the sulfur chemistry. Lastly, we sum up our conclusions in Sect.~\ref{sec:conclusion}. 

\begin{table*}[htb!]
    \centering
    \caption{Table containing the relevant spectroscopy line properties of the targeted \ce{CH3OH} transitions, as well as the beam size, beam position angle (PA) and channel rms inside the imaged line cubes.}
    \addtolength{\leftskip} {-2cm} \addtolength{\rightskip}{-2cm}
    \begin{tabular}{ccccccccc}
        \hline
        Transition & $\nu$ & $\log_{10}A_{ul}$  & $E_{u}$ & $g_{u}$ & Program & Beam size & Beam PA & channel rms \\
        &  [GHz] & [$s^{-1}$] & $[\mathrm{K}]$ & & & [$''$] &[°] & [mJy/beam] \\
        \hline
        $3_{(-1,2)}-2_{(-0,2)}$ E & 310.1930 & -4.056 & 34.98 & 28 & 2022.1.00905.S & $0.83 \times 0.65$ & -47.83 & 0.76 \\
        $6_{(1,5)}-6_{(0,6)}$ A* & 311.8526 & -3.466 & 63.71 & 52 & 2017.1.01178.S & $0.39 \times 0.35$ & -28.57 & 2.04 \\
        $6_{(1,5)}-6_{(0,6)}$ A & 311.8526 & -3.466 & 63.71 & 52 & 2022.1.00905.S & $0.82 \times 0.65$ & -47.92 &0.70 \\
        $7_{(1,7)}-6_{(1,6)}$ A & 335.5820 & -3.789 & 78.97 & 60 & 2017.1.01178.S & $0.34 \times 0.32$ & -24.10 & 2.98 \\
        $7_{(1,7)}-6_{(1,6)}$ A** & 335.5820 & -3.789 & 78.97 & 60 & 2019.1.00393.S & $0.64 \times 0.39$ & -49.53 & 2.53 \\
        $7_{(1,6)}-7_{(0,7)}$ A* & 314.8585 & -3.456 & 80.09 & 60  & 2017.1.01178.S & $0.38 \times 0.35$ & -28.98 & 2.10 \\
        $7_{(-1,6)}-6_{(-2,5)}$ E & 313.5968 & -4.327 & 86.05 & 60 & 2022.1.00905.S & $0.81 \times 0.66$ & -49.07 & 0.70 \\
        $9_{(1,8)}-9_{(0,9)}$ A* & 322.2395 & -3.433 & 119.88 & 76 & 2017.1.01178.S & $0.34 \times 0.32$ & -23.91 & 4.85 \\
        $9_{(1,8)}-9_{(0,9)}$ A & 322.2395 & -3.433 & 119.88 & 76 & 2022.1.00905.S & $0.78 \times 0.64$ & -46.99 & 1.60 \\
        $10_{(-0,10)}-9_{(-1,8)}$ E* & 314.3511 & -4.108 & 140.60 & 84 & 2017.1.01178.S & $0.39 \times 0.35$ & -28.87 &  2.08\\
        $12_{(1,11)}-12_{(0,12)}$ A** & 336.8651 & -3.390 & 197.08 & 100 & 2019.1.00393.S & $0.63 \times 0.39$ & -49.56 & 2.27 \\
        Stacked lines* & / & / & / & / & 2017.1.01178.S & $0.38 \times 0.35$ & -28.25 & 1.18 \\
        Stacked lines** & / & / & / & / & 2019.1.00393.S & $0.63 \times 0.39$ & -49.56 & 1.61 \\
        \hline
    \end{tabular}      
    \tablefoot{\small{{The spectroscopy line properties are from from the CDMS database (\citealt{CDMS_1, CDMS_2, CDMS_3, CDMS_meth}).  All the methanol lines were imaged with a 1 km s$^{-1}$ channel spacing. The channel rms is computed after the 1 km s$^{-1}$ binning. $^{(*)}$ Lines covered in program 2017.1.01178.S used to create the Stacked lines* entry. $^{(**)}$ Same but then for program 2019.1.00393.S.}}}
\label{tab:meth_lines}
\end{table*}

\section{Observations} \label{sec:ALMA_archive}

\subsection{Criteria for dataset selection}\label{sec:selec_criteria}

Due to its asymmetric three dimensional geometric shape, the methanol molecule has access to a great number of possible rotational transitions at temperatures close to its desorption temperature of $\sim$ 120 K. This means that almost every ALMA archival observation of HL Tau covers at least one methanol transition. Therefore, we selected only the datasets that cover \ce{CH_{3}OH} transitions that are predicted to be bright, assuming LTE and optically thin emission. \par 

Indeed, under these assumptions, it is possible to relate $N^{\rm thin}_{\rm u}$, the column density of the atoms/molecules excited in the upper energy level $E_{\rm u}$, to $S_{\nu}$, the flux density integrated over a velocity range $\Delta V$:

\begin{equation}  \label{eq:Line_N_thin}
    N^{\rm thin}_{\rm u} = \frac{4\pi}{A_{\rm ul}hc} \frac{S_{\nu}\Delta V}{\Omega}.
\end{equation}

In the formula above, $c$ is velocity of light, $h$ is the Plank constant, $\nu$ is the frequency, $A_{\rm ul}$ is the Einstein coefficient of the transition, and $\Omega$ is the solid angle subtended by the emitting area of the source. Now, applying the natural logarithm to both sides of Boltzmann equation after inserting equation \ref{eq:Line_N_thin}, it is possible to find that 

\begin{equation} \label{eq:Rot_diagram}
    \ln \left (\frac{N^{\rm thin}_{\rm u}}{g_{\rm u}} \right) =  \ln N_{\rm T} -\ln{Q(T_{\rm ex})}  -\frac{E_{\rm u}}{kT_{\rm ex}}.
\end{equation}

The expression above links the excitation temperature $T_{\rm ex}$ and the average column density of the emitting species $N_{\rm T}$ to the ratio between $N^{\rm thin}_{\rm u}$ and the degeneracy of the upper energy level $g_{\rm u}$. $Q(T_{\rm ex})$ is the partition function of the considered atom/molecule and $k$ is the Boltzmann constant. Thus, it is possible to estimate the expected line flux $S_{\nu} \Delta V$ through the following expression, derivable from combining together eq.~\ref{eq:Line_N_thin} and eq.~\ref{eq:Rot_diagram}:

\begin{equation} \label{eq:Alminer_formula}
\begin{split}
{S_{\nu}\Delta V} &=  \frac{hc \Omega}{4\pi} \frac{A_{\rm ul} g_{\rm u}}{Q(T_{\rm ex})} N_{\rm T} \exp  \left(-\frac{E_{\rm u}}{kT_{\rm ex}} \right) 
\\ 
&\propto \frac{A_{\rm ul} g_{\rm u}}{Q(T_{\rm ex})} \exp  \left(-\frac{E_{\rm u}}{kT_{\rm ex}} \right).
\end{split}
\end{equation}

Applying equation \ref{eq:Alminer_formula} to the \ce{CH_{3}OH} line list from the CDMS database (\citealt{CDMS_1, CDMS_2, CDMS_3, CDMS_meth}), we ranked all the methanol rotational transitions within the 70-700 GHz frequency interval, according to the expected line strength for an assumed $T_{\rm ex}=150$ K. The latter was chosen as a representative value of the $T_{\rm ex}$ of methanol based on the one computed for the thermally desorbed \ce{CH3OH} reservoir in the HD 100546 disk \citep{Evans_2025} and on the temperature derived from the two water transitions with the lowest $E_{\rm u}$ in HL~Tau \citep{Facchini_2024}. \par

In order to pinpoint the best observational programs to search for methanol, taking into account at the same time both the intrinsic properties of the molecular transitions and the technical features of the observation, we employed built-in methods from the \verb+Alminer+ library \citep{Alminer_2023} to cross-correlate the ranked \ce{CH3OH} line list with the ALMA archive. We gathered at first every HL~Tau observation that was covering at least one methanol transition, rejecting those with coarse spectral (>5 km s$^{-1}$) or angular ($> 3''$) resolution. These cut-offs were roughly chosen in order to restrict our query only to Frequency Division Mode (FDM) observations conducted with the 12-m array. Then, the list of these \ce{CH3OH} lines was ranked in decreasing order based on the ratio of the expected line flux obtained through eq. \ref{eq:Alminer_formula} and the line sensitivity of the observations. We selected all the datasets covering at least one methanol transition with an estimated ratio within one third of the maximum value found for the lines inside the 70-700 GHz interval. Table~\ref{tab:meth_lines} summarizes the information on the methanol lines, identified by their quantum numbers, that we imaged and subsequently analyzed for this project. Three lines ($7_{(1,6)}-7_{(0,7)}$ A, $9_{(1,8)}-9_{(0,9)}$ A, $6_{(1,5)}-6_{(0,6)}$ A) were observed by more than one program.

\subsection{Observations properties and line imaging}\label{sec:imaging}

The dataset covering the selected \ce{CH3OH} lines belong to Band 7 ALMA observations from three programs: 2017.1.01178.S (PI: Humphreys, E.), 2022.1.00905.S (PI: Facchini, S.) and 2019.1.00393.S (PI: Zhang, K.). Detailed information about the observing set, calibrators and spectral settings for each of these programs are presented in App.~\ref{sec:Appendix_A}. In App.~\ref{sec:Appendix_B} we discuss the bandpass calibration of the observations from program 2022.1.00905.S, which are not thermal-noise limited but spectral dynamical range limited. \par 
The transitions in Table \ref{tab:meth_lines} were imaged employing the processing software CASA, version \verb+CASA 6.5.4-9+ \citep{CASA_2022}. We started with subtracting the continuum emission of the HL Tau disk using the CASA task \verb+uvcontsub+ fitting a first order polynomial, after flagging the channels covering bright lines. We then imaged every spectral window using \verb+tclean+ and setting the parameter \verb+niter+ to one. Due to the absence of any emission ascribable to methanol, we reconstructed the dirty (\verb+niter+ = 0) cubes of all the methanol lines in our sample imaging one hundred channels around the frequency of each \ce{CH3OH} transition. We chose a 1~km~s$^{-1}$ channel spacing, close to the maximum velocity resolution allowed by the 976 kHz channel spectral windows (see App. \ref{sec:Appendix_A}),  and we did not apply the primary beam correction. We imaged the lines with Briggs weighting \citep{Briggs_1995}, setting the \verb+robust+ parameter at 2, in order to boost the sensitivity of the resulting data cubes at the expense of a lower angular resolution. As the B7 observations of program 2017.1.01178.S where taken at a very high angular resolution of $0.04''-0.08''$, limiting the sensitivity to extended emission, we applied a $0.3''$ $uv$-taper. \par
Both the beam axes and the channel rms of the imaged line cubes are reported in Table~\ref{tab:meth_lines}. The noise has been evaluated with the \verb+estimate_RMS+ of the \verb+GoFish+ package \citep{GoFish}, which computes the line cube rms in an user-defined annular region, and using the first 10 and last 10 channels, chosen as they are line-free after visual inspection from the data. We set the inner and outer radius at, respectively, 3$''$ and 8$''$, in order to avoid including the emission coming from the disk. \par

\subsection{Line stacking}\label{sec:stacking}

We also stacked together in the $uv$-plane four \ce{CH3OH} lines from the 2017.1.01178.S program ($6_{(1,5)}-6_{(0,6)}$ A, $E_{\rm u}$ = 63.71 K; $7_{(1,6)}-7_{(0,7)}$ A, $E_{\rm u}$ = 80.09 K; $9_{(1,8)}-9_{(0,9)}$ A, $E_{\rm u}$ = 119.88~K; $10_{(-0,10)}-9_{(-1,8)}$ E, $E_{\rm u}$ = 140.60 K). Since it is not possible to stack multiple lines in the frequency domain, the
procedure consists first in regredding the data with the CASA task \verb+cvel2+ in velocity mode, assigning the systemic velocity of the system to the frequency of each line. Afterwards, we concatenated the datasets into one and imaged it with the same parameters described above, obtaining a channel rms of 1.18 mJy/beam. Since the two \ce{CH_{3}OH} transitions in program 2019.1.00393.S ($7_{(1,7)}-6_{(1,6)}$~A, $E_{\rm u}$ = 78.97~K; $12_{(1,11)}-12_{(0,12)}$~A, $E_{\rm u}$ = 197.08~K) belonged to the same spectral window, we also stacked only these two lines together. The resulting data cube has an improved channel rms of 1.60~mJy/beam compared to individual values >~2.2~mJy/beam. \par

\begin{table*}[htb!]
    \centering
    \caption{$3\sigma$ upper limits on the disk integrated \ce{CH3OH} flux and \ce{CH3OH} column density for various assumed excitation temperatures.}\captionsetup{justification=centering}
    \begin{tabular}{cccccccc}
        \hline 
        Transition & $\nu$ & Program & spectrum rms & $3\sigma_{S_{\nu} \Delta V}$ & $\mathrm{log_{10}} \left( N_{\ce{CH_{3}OH}} \right)$ & $\mathrm{log_{10}} \left(N_{\ce{CH_{3}OH}} \right)$ & $\mathrm{log_{10}} \left(N_{\ce{CH_{3}OH}} \right)$ \\
        & [GHz] & & [mJy] & [mJy km s$^{-1}$] & at 100 K &  at 168 K$^{(a)}$&  at 200 K  \\
        \hline
        $3_{(-1,2)}-2_{(-0,2)}$ E & 310.1930 & 2022.1.00905.S & 1.0 & 10.4 & 15.6 & 16.0 & 16.1 \\
        $6_{(1,5)}-6_{(0,6)}$ A & 311.8526 & 2017.1.01178.S & 6.1 & 63.6 & 15.7 & 16.0 & 16.1 \\
        & & 2022.1.00905.S & 0.9 & 9.8 & 14.9 & 15.1 & 15.3 \\
        $7_{(1,7)}-6_{(1,6)}$ A & 335.5820 & 2017.1.01178.S & 7.7 & 80.5 & 16.1 & 16.4 & 16.5 \\
        & & 2019.1.00393.S & 6.0 & 62.6 & 16.0 & 16.3 & 16.4 \\
        $7_{(1,6)}-7_{(0,7)}$ A & 314.8585 & 2017.1.01178.S & 6.6 & 69.0 & 15.7 & 16.0 & 16.1 \\
        $7_{(-1,6)}-6_{(-2,5)}$ E & 313.5968 & 2022.1.00905.S & 1.0 & 10.5 & 15.8 & 16.0 & 16.1 \\
        $9_{(1,8)}-9_{(0,9)}$ A & 322.2395 & 2017.1.01178.S & 12.8 & 133.9 & 16.0 & 16.2 & 16.3 \\
        & & 2022.1.00905.S & 2.3 & 23.8 & 15.3 & 15.5 & 15.6 \\
        $10_{(-0,10)}-9_{(-1,8)}$ E & 314.3511 & 2017.1.01178.S & 6.2 & 64.2 & 16.4 & 16.6 & 16.7 \\
        $12_{(1,11)}-12_{(0,12)}$ A & 336.8651 & 2019.1.00393.S & 5.4  & 56.7 & 15.8 & 15.9 & 16.0 \\
        \hline
    \end{tabular}
    \tablefoot{\small{In order, the column entries are: the quantum number of the imaged methanol transition, its frequency, the observational program covering the \ce{CH3OH} line, the rms in the spectrum, the 3$\sigma$ upper limit on the integrated line flux assuming a line width of 10 km s$^{-1}$ and the  3$\sigma$ upper limit on the methanol column density $N_{\ce{CH3OH}}$ for an excitation temperature of 100, 168 and 200 K, respectively. $^{(a)}$ The temperature of 168 K, residing inside the likely interval of 100-200 K, is the rotational temperature found by \cite{Facchini_2024} for the two water lines with the lowest $E_{\rm u}$.}}
    \label{tab:meth_col_dens}
\end{table*}

\begin{figure*} [htb!]
\centering
\includegraphics[width = 1. \linewidth]{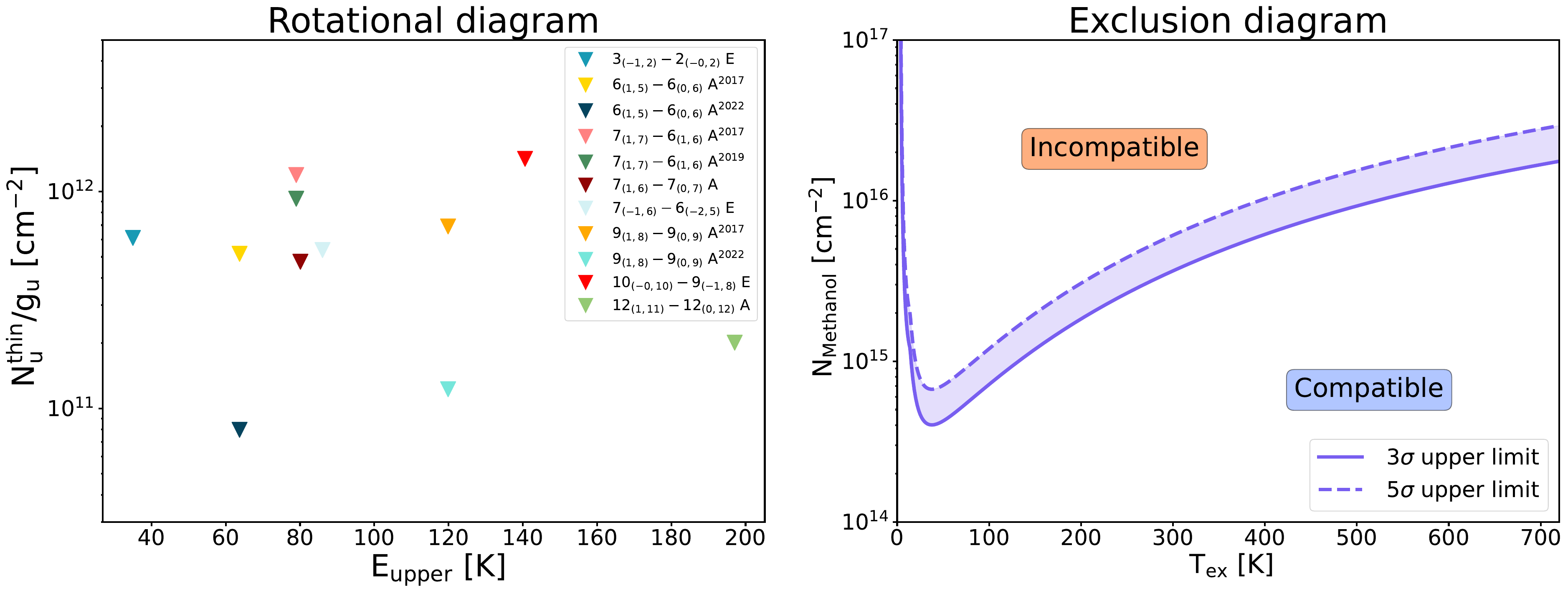}
\caption{\small{The left panel shows a rotational diagram built from the methanol non-detections. In the legend, each transition is named after its quantum numbers. For the lines that are covered in multiple programs, the year of the program is also provided. The right panel displays an exclusion diagram highlighting the values of excitation temperature and total column density compatible with all the $3\sigma$ (presented in the left panel) and $5\sigma$ upper limits on the total \ce{CH_{3}OH} column density.}}
\label{fig:Rot_diagram}
\end{figure*}


\section{Results}\label{sec:result}
For each data cube covering the targeted methanol lines, we extracted the spectra from a circular area centered on HL Tau with a radius of 0.7$''$, employing available tools from the \verb+GoFish+ library. This mask corresponds to the spatial location of the \ce{H2O} emission in this disk \citep{Facchini_2024, Leemker_2025_A&A}. In App.~\ref{sec:Appendix_F}, Fig. \ref{fig:H_spectra}, \ref{fig:S_spectra}, \ref{fig:Z_spectra} display, respectively, the spectra from the lines covered in the programs 2017.1.01178.S, 2022.1.00905.S, and 2019.1.00393.S. Methanol emission was not detected in any of the 13 analyzed spectra. From this point onward, we will not consider anymore the two data cubes obtained by stacking together the lines from program 2017.1.01178.S and 2019.1.00393.S, due to the different upper energy levels of the combined transitions, which differ more than 60 K. \par

Following the methods of \cite{vantHoff+2020}, when a line is not detected, it is still  possible to derive the 3$\sigma$ upper limit on $N^{\rm thin}_{\rm u}$ from the spectrum rms ($\sigma_{\rm spectrum}$), replacing $S_{\nu} \Delta V$ in eq. \ref{eq:Line_N_thin} with the the 3$\sigma$ upper limit on the integrated line flux: 

\begin{equation}\label{eq:error}
    3\sigma_{S_{\nu} \Delta V} = 3 \times 1.1 \sqrt{\Delta V \delta V} \times \sigma_{\rm spectrum},
\end{equation}

where $\Delta V$ is now the expected line width of the line and $\delta V$ is the spectral resolution, both in km s$^{-1}$. The 1.1 multiplicative factor accounts for the 10\% uncertainty on the flux calibration of ALMA Band 7 (B7; \citealt{, Francis+2020}). In our analysis, we estimate $\sigma_{\rm spectrum}$ as the mean of the rms in the spectra extracted from 50 random, non-overlapping, emission-free circular regions of radius 0.7$''$. We report in Table~\ref{tab:meth_col_dens} the spectral noise estimated following this method. \par

To compute the $3\sigma$ upper limits on $N_{\rm u}^{\rm thin}$ through eq.~\ref{eq:Line_N_thin} and eq.~\ref{eq:error}, we assumed a line width $\Delta V$ of 10 km s$^{-1}$, corresponding to the approximate width of the detected water line with the lowest $E_{\rm u}$, and we also assumed that the emitting area $\Omega$ is confined inside the 17 au snowline constrained by \cite{Facchini_2024}. The latter is motivated by the consideration that, if \ce{CH3OH} is thermally desorbed from the surface of dust grains in the disk, the main gas-phase methanol reservoir should be co-spatial to the water one. The left panel of Figure~\ref{fig:Rot_diagram} shows how the upper limits $N_{\rm u}^{\rm thin}$, divided by $g_{\rm u}$, vary with the $E_{\rm u}$ of each methanol transition in the rotational diagram. We note that a factor of 2 difference in the assumed $\Delta V$ would translate into a 40\% difference in this upper limit. In contrast, an uncertainty of a factor of two in the radius of the circular emitting area $\Omega$ would result in a factor of four difference in $N_{\rm u}^{\rm thin}$, which cancels out when computing the column density ratio with other molecules, as we assume that both molecules emit from the same region.\par 

As none of the imaged \ce{CH3OH} lines are detected, upper limits on the \ce{CH3OH} column density $N_{\ce{CH3OH}}$ are derived following the procedure outlined in \citet{Rot_Diagram} and by assuming a $T_{\rm ex}$ following eq.~\ref{eq:Rot_diagram}. The right panel of Fig.~\ref{fig:Rot_diagram} shows an exclusion diagram displaying the regions in the ($T_{\rm ex}$, $N_{\ce{CH3OH}}$) parameter space compatible with the derived upper limits on the methanol column density. To create this figure, we combined the 3$\sigma$ and 5$\sigma$ upper limits on $N^{\rm thin}_{\rm u}$, varying the assumed excitation temperature from a non-physical $T_{\rm ex}$ of 0 K to 700 K, close to the maximum possible value of the rotational temperature found by \cite{Facchini_2024} for the three detected water lines in HL Tau. For temperatures below 20 K, the exclusion diagram does not give any meaningful constraint on the methanol column density, while at the temperature of $\sim$ 38 K, it returns the lowest possible $3\sigma$ upper limit ($N_{\ce{CH3OH}}$ < $4.0\times10^{14} \ \mathrm{cm^{-2}}$).

Table~\ref{tab:meth_col_dens} reports for every \ce{CH_{3}OH} line the $3 \sigma$ upper limits on $N_{\rm T}$ assuming three likely excitation temperatures for thermalized methanol emission. Indeed, it is to be expected that the true \ce{CH3OH} $T_{\rm ex}$ resides between the 100-200 K interval, based on the binding energy of \ce{CH_{3}OH} to water ice (\citealt{Penteado_2017, Minissale_2022}) and on the methanol excitation temperature measured via rotational diagram analysis in the HD 100453 and HD 100546 protoplanetary disks (\citealt{Booth_2025a, Evans_2025}). Within this temperature interval, the $6_{(1,5)}-6_{(0,6)}$ A line from program 2022.1.00905.S gives the most stringent constraint on the methanol column density in the HL Tau disk, which range between  $7.2\times10^{14} \ \mathrm{cm^{-2}}$ ($T_{\rm ex}$=100 K) -  $1.8\times10^{15} \ \mathrm{cm^{-2}}$ ($T_{\rm ex}$=200 K). \par 
These values are between one and two orders of magnitude higher than the upper limit measured in the HL Tau disk by \cite{Garufi_2021}, using the $5_{(0,5)}-4_{(0,4)}$ A line at 241.79 GHz. This methanol transition is slightly colder ($\mathrm{log_{10}}A_{\rm ul}$ = -4.22 and $E_{\rm u}$ = 34.81 K) than the $6_{(1,5)}-6_{(0,6)}$ A line ($\mathrm{log_{10}}A_{\rm ul}$ = -3.47 and $E_{\rm u}$ = 63.71 K). The range of column density upper limits (1.9-7.5$\times10^{13} \ \mathrm{cm^{-2}}$) was estimated assuming an excitation temperature of 20 K and 100 K, while an annular region with inner and outer radius of, respectively, 55 and 250 au was assumed as the methanol emitting area. Since $N_{\ce{CH3OH}}$ is inversely proportional to $\Omega$, the difference between our values and \cite{Garufi_2021} ones can be explained by our assumed emitting area (a circular region of radius 17 au), which is $\sim$ 200 times smaller.


\begin{figure*} [htb!]
\centering
\includegraphics[width = 1 \linewidth]{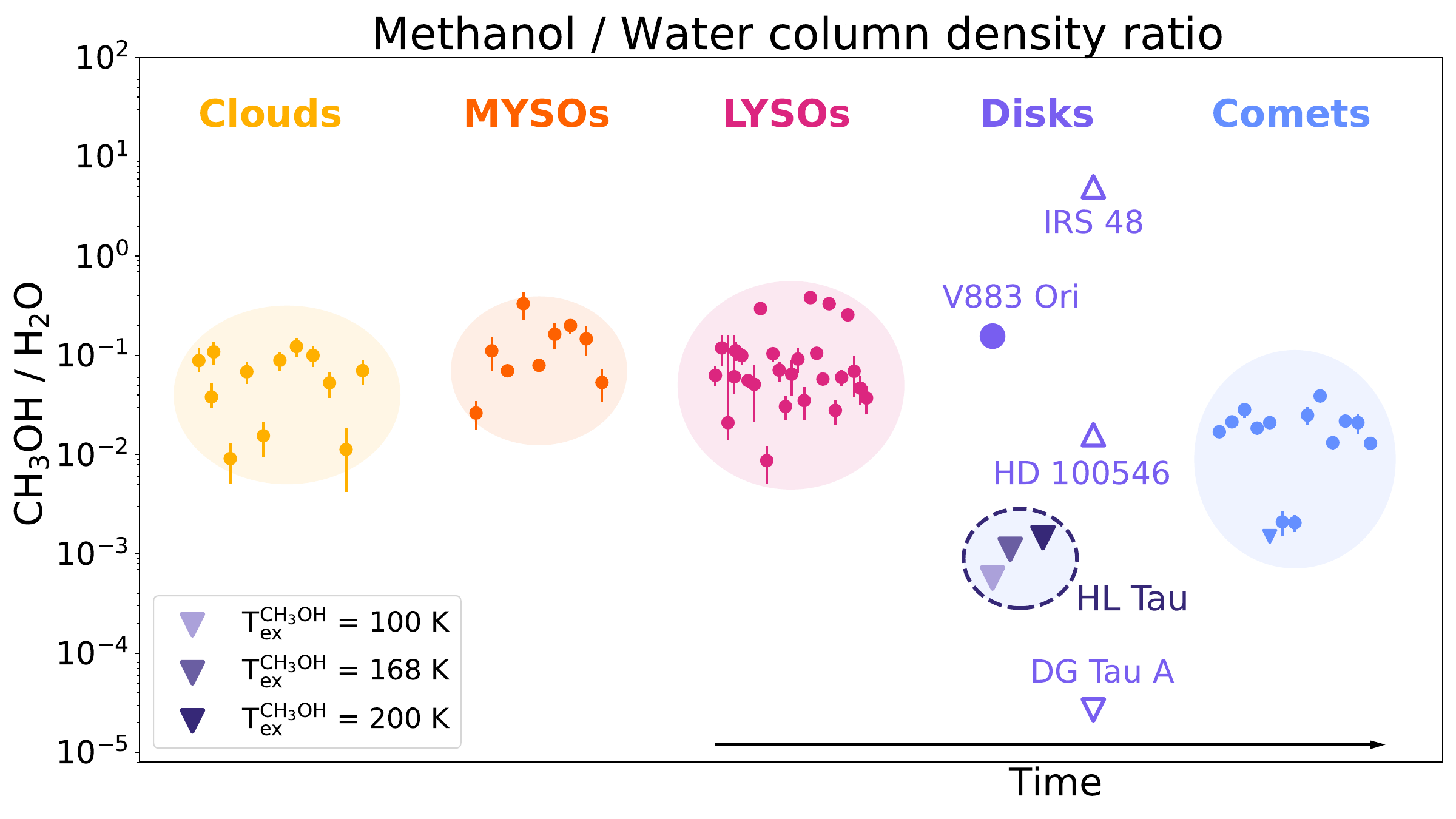}
\caption{\small{\ce{CH3OH}/\ce{H2O} column density ratio found in the literature for a wide range of clouds, YSOs and Solar System comets (see Sect~\ref{sec:meth_to_water} for the references). The values for molecular clouds, LYSOs and MYSOs reflect the ice composition, while the ratio was measured in the gas-phase for protoplanetary disks and comets. We used open symbols for the protoplanetary disks in which the upper/lower limit on the methanol-to-water column density ratio was estimated combining measurements taken at different wavelength ranges. The upward triangles for the HD~100546 and the IRS~48 disks indicate a lower limit, whereas the downward triangles for DG~Tau~A and HL~Tau indicate a upper limit.} The values for HL Tau are taken dividing the $3\sigma$ upper limit on the $N_{\ce{CH3OH}}$ for the three different assumed excitation temperature in Table \ref{tab:meth_col_dens} for the water column density measured by \cite{Facchini_2024}.}
\label{fig:Met_to_water}
\end{figure*}

\section{Discussion}\label{sec:discussion}

\subsection{Methanol-to-water column density ratio}\label{sec:meth_to_water}

From the collapse of a molecular cloud to the formation of a protoplanetary disk, the methanol budget remains closely tied to that of water, owing to their similar volatilities. Consequently, constraining the change of the methanol-to-water column density ratio, both in the ices and in the gas-phase, provides valuable insights into the degree of chemical processing that these two molecules undergo during the evolution of a YSO. \par 
Figure \ref{fig:Met_to_water} presents the \ce{CH3OH}/\ce{H2O} column density ratio measured for several clouds, Young Stellar Objects, namely low- and high-mass YSOs (called, respectively, as LYSOs and MYSOs), protoplanetary disks and Solar System comets, which are believed to be the leftovers of the planet formation mechanisms that interested our planetary system more than 4 Gyr ago \citep{OOrt_1950}. This plot combines measurements taken with different telescopes and targeting molecules in both gas and ice phase. The values for the molecular clouds, LYSOs and MYSOs, measured observing the ice absorption feature in the infrared (IR) spectra of background stars or of the protostars themselves, are from \cite{Whittet_2011} and references therein. We also used the recent \ce{CH3OH}/\ce{H2O} ratios for two clouds and four LYSOs observed with JWST \citep{McClure_2023, Chen_2024, Rocha_2024, Rocha_2025}. At least for LYSOs, radio-interferometric observations of gas-phase \ce{CH3OH} and H$_{2}^{18}$O yield column density ratios consistent with those found in the ices \citep{Jensen_21, Jacobsen_19, Manigand_2020, Persson_2013, Okoda_24}. The values for the comets, obtained with IR spectroscopy, are taken instead from the references listed in the review by \cite{Mumma_2011}, while the specific ratio for the 67P/Churyumov-Gerasimenko comet comes from the ROSINA mass-spectrometer on board on the Rosetta spacecraft, during an outgassing event of the comet near its perihelion \citep{Rubin_2019}. Lastly, the values for the protoplanetary disks are derived by combining gas-phase methanol column densities from the literature (see Table~\ref{tab:meth_in_disk}; except for HL~Tau) with water vapor column density measurements or upper limits also reported in the literature. The water column density upper limits for the HD~100546 \citep{Pirovano_2022} and IRS~48 \citep{Leemker_2023} disks are based on \textit{Herschel} Far-IR data, whereas the values for V883~Ori \citep{Tobin_2023} and HL~Tau \citep{Facchini_2024, Leemker_2025_A&A} have been measured with ALMA at sub-millimeter wavelengths. Water has also been detected in the DG~Tau~A protoplanetary disk in the Far-IR with \textit{Herschel} \citep{Podio_2013}. The upper limit on the methanol-to-water ratio for this source was obtained by converting the total water vapor mass estimated from thermochemical modeling into a water column density and combining it with the methanol column density upper limit from \cite{Podio_2019}, after scaling to the same emitting area. In the plot, we report three values for HL~Tau, derived assuming a methanol excitation temperature of 100, 168, and 200~K and dividing by the water column density retrieved by \cite{Facchini_2024} for an emitting area of $\sim$17~au. \par

In the light of the above, assuming for all the sources optically thin emission, the plot in figure \ref{fig:Met_to_water} shows the variation of the \ce{CH3OH}/\ce{H2O} ratio over time. The ratio is approximately constant for molecular clouds and Class 0/I YSOs still embedded in their collapsing envelope, while it is on average a factor of 5 lower for Solar System comets. \par 

Different is the case for protoplanetary disks: the only five sources with measurements or upper limits display a large scatter, spanning over five orders of magnitude. The V883~Ori disk is the only one with a methanol-to-water ratio consistent with clouds, LYSOs and MYSOs. In this source, the central protostar is undergoing an accretion burst, effectively shifting the water snowline in the outer disk. Both the emission of the freshly sublimated water and methanol originates from an annular region, due to the highly optically thick dust residing in the center of the disk and, possibly, due to also a depletion in the COM abundances in the inner part \citep{Lee_2019, Leemker_2021, Tobin_2023, Yamato+24}. The similarity between the gas-phase methanol-to-water column density ratio measured in the disk of V883~Ori and the one observed in the ices of other YSOs further establishes that the bulk of the icy material is inherited by disks from the interstellar medium, without undergoing extensive chemical reprocessing \citep{Booth_2021, Tobin_2023, Leemker_2025_Nature}. \par

In contrast, the disk around the IRS~48 stars displays a~$\sim$~two order of magnitude higher ratio, hinting a potential different chemical evolution in respect to V883~Ori. The high \ce{CH3OH}/\ce{H2O} ratio ($\gtrsim$ 5; \citealt{Leemker_2023}) for the IRS~48 system may be partially explained by the fact the methanol is sub-thermally excited, consequently invalidating the LTE assumption used to derive the \ce{CH3OH} column density \citep{Temmink_2024}. Moreover, IRS~48 is a peculiar source, hosting a big asymmetrical, crescently shaped, dust trap located at 60-80 au from the central star \citep{VanderMarel_2013}. Inside the emission of several COMs and oxygen bearing molecules related to the thermal desorption of ice has been detected \citep{VanDerMarel_2021, Brunken_2022, Leemker_2023, Booth_2024b}. Therefore, the high \ce{CH3OH}/\ce{H2O} ratio can be due to a different chemical composition of the ices in the trap, due to other effects altering the ratio once the two species sublimates into gas-phase or due to the different wavelength ranges of the probed water and methanol emission lines \citep{VanDerMarel_2021, Leemker_2023}. Similarly, the upper and lower limit for the DG~Tau~A and HD~100546 disks are derived, respectively, from \ce{H2O} observations with \textit{Herschel} in the far-IR and \ce{CH3OH} observations with ALMA in the~(sub-)mm. For these three disks, the water column density has been computed from thermochemical models. \par 

Conversely, the \ce{CH3OH}/\ce{H2O} ratio in the HL~Tau disk is derived from radio-interferometric observations taken at approximately the same wavelength with ALMA, similar to that in the V883~Ori disk. This allows for a direct comparison between the disks in a similar part of the electromagnetic spectrum. Combining together the upper limit on the methanol column density reported in Table \ref{tab:meth_col_dens} with the water column density measured by \cite{Facchini_2024}, we found a methanol-to-water column density ratio upper limit for HL Tau (< $0.55 \times 10^{-3}$ at 100 K and < $1.4 \times 10^{-3}$ at 200 K) which is one order of magnitude lower than the one measured for other YSOs and for most of the Solar System Comets. This ratio can be explained by radiative transfer effects or by a chemical difference. \par

Focusing on the former, we neglect the wavelength dependency of the dust opacity, since the methanol and water column densities were computed from lines with roughly the same frequency. Referring to the discussion in \cite{Isella_2016} and \cite{Weaver_2018}, we consider the following possibilities: 

\begin{itemize}
    \item Both the \ce{CH3OH} and \ce{H2O} emission are optically thin or become optically thick below $\tau_{\rm dust}$ = 1, while the dust is optically thick: the only line photons that escape are emitted in the layer of the disk where $\tau_{\rm dust} \lesssim 1$. As the same disk region is probed, no major changes in the \ce{CH3OH}/\ce{H2O} column density ratio are expected due to radiative transfer effects in this case.
    \item Water is optically thick above the surface where the continuum becomes optically thick, while methanol is optically thin: if that is the case, the water column density is underestimated not only due to part of the \ce{H2O} reservoir being hidden by optically thick dust, but also due to the optical depth of the line itself and due to continuum oversubtraction. This effect would lead to an even lower methanol-to-water column density ratio.
    \item The water and methanol emission are optically thick above the $\tau_{\rm dust}$ = 1 surface: the values for the column densities are not representative of the true \ce{CH3OH} and \ce{H2O} abundance. To test wether the \ce{CH3OH} lines could be optically thick, we estimated the maximum emitting area that an optically thick methanol reservoir could have while still being undetected. Given the channel rms in Kelvin of each line cube and assuming a methanol excitation temperature of $\sim$120 K, the 3$\sigma$ upper limit on the maximum methanol emitting area $A_{\ce{CH3OH}}$ is computed as $A_{\ce{CH3OH}} = (3 \cdot \sigma_{\rm channel \ rms} / T_{ex}) \cdot A_{\rm beam}$, where $\rm A_{\rm beam}$ is the beam area. The tightest constrain on $A_{\ce{CH3OH}}$ is given by the $6_{(1,5)}-6_{(0,6)}$ A line from program 2022.1.00905.S: to be both optically thick and undetected at the same time, the methanol emission should come from circular region with a radius $\lesssim$ 1.2~au. Since this radius is a factor of four less than the estimated location of the water snowline in the disk ($\sim$~5~au), it is unlikely that the methanol emission is optically thick \citep{Guerra-Alvarado+24}. 
    
    \item If water is masing, the methanol-to-water column density ratio in HL Tau would be higher. This could occur if the high \ce{H2O} energy levels are overpopulated in respect of the LTE scenario and, as a consequence, the water column density derived by \cite{Facchini_2024} would overestimate the actual \ce{H2O} column density. While masing cannot be completely excluded, \cite{Leemker_2025_A&A} recently showed that water is likely at most weakly masing in the HL~Tau disk.
\end{itemize}

Instead, if we assume that the stringent upper limit on the methanol-to-water ratio is representative of the true methanol and water bulk abundance inside the disk, we are likely tracing a chemical effect and/or a methanol depletion. A decrease of this ratio over the evolution of a Young Stellar Object is expected if the molecules desorb into the gas-phase, after the bulk of the ice reservoir is inherited from the earliest phases of star and planet formation \citep{Booth_2021, Tobin_2023, Leemker_2025_Nature}. Indeed, methanol is assembled efficiently just onto the surface of ISM dust grains and is dissociated two times faster than water, which, conversely, can also be formed in the gas if temperature exceeds 300 K. Taking into account the combination of these mechanisms, the gas-phase \ce{CH3OH}/\ce{H2O} ratio is expected to stay constant or decrease with time. A similar effect would occur if the water emission in HL Tau is arising from a wind associated to the CO-traced outflow \citep{Baciotti+25}: if that would be the case, \ce{H2O} and \ce{CH3OH} would get rapidly dissociated but, due to the higher temperatures reached by the gas, \ce{H2O} would reformed directly into gas-phase. \par
To sum up, we have listed the radiative transfer and chemical effects that could explain the low upper limit on the methanol-to-water column density ratio found for the HL Tau protoplanetary disk. Nevertheless, other explanations such as an intrinsically low \ce{CH3OH} abundance cannot be ruled out.

\begin{figure*} [htb!]
\centering
\includegraphics[width = 1. \linewidth]{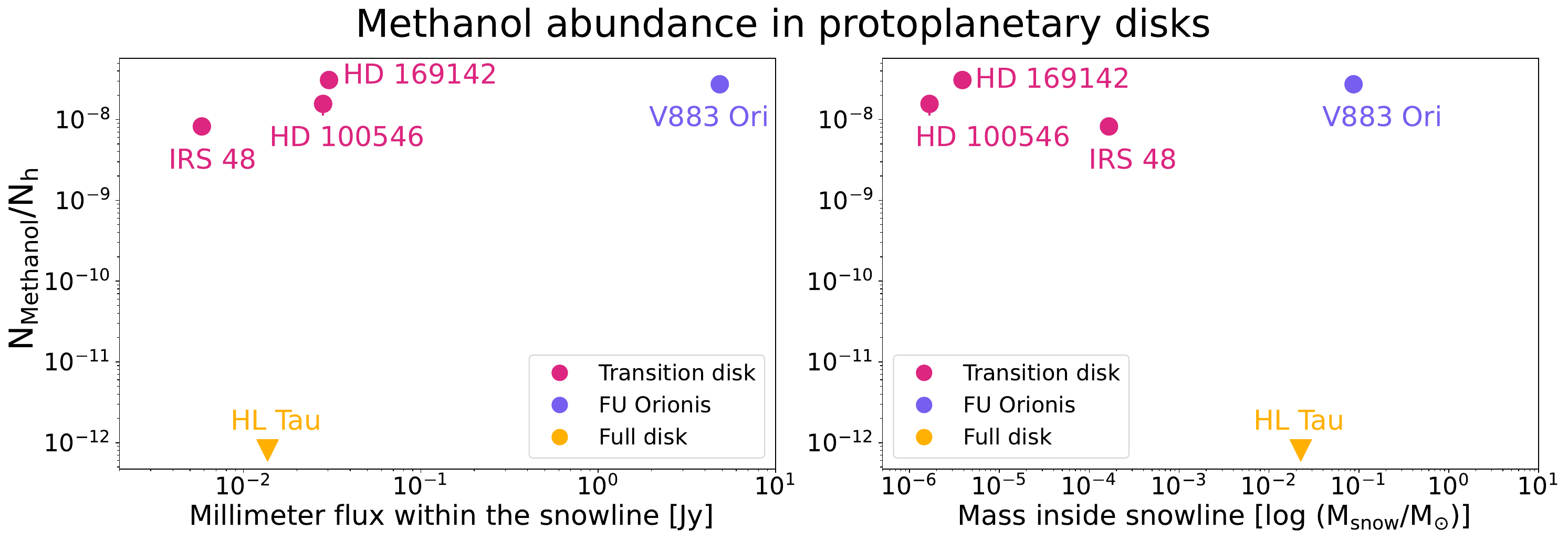}
\caption{\small{Ratio between the methanol and the average hydrogen column density versus the Band 7 continuum flux integrated within the methanol snowline (left panel) and versus the mass enclosed inside $r_{snow}$ (right panel). For the HL Tau disk, the $3\sigma$ upper limit on the $N_{\ce{CH3OH}}$ is presented, assuming excitation temperature of 168~K. The low methanol abundance upper limit in the HL~Tau disk is likely motivated by the different dust optical depth and by the different stellar luminosity in respect of the other disks in our sample.}}
\label{fig:meth_corr}
\end{figure*}

\subsection{Methanol budget in protoplanetary disks and the effect of optically thick dust}

Comparing the stringent upper limit on the methanol column density in the HL Tau disk with the one measured in other protoplanetary disks can shed light on the reason why \ce{CH3OH} was not detected in this source. In addition to the disks with constrains on both the methanol and the water column densities, \ce{CH3OH} has also been observed in the TW~Hya, HD~169142, HD~100453 and CQ~Tau disks, all of which are transition disks \citep{Walsh_2016, Booth_2021, Booth_2025a, Booth_2025b}. Table~\ref{tab:meth_in_disk} summarizes the mass and the luminosity of the star hosted by the disks with methanol detection, as well the \ce{CH3OH} column density values corresponding to the given excitation temperatures. We also report the stellar parameters of HL Tau and the \ce{CH3OH} column density upper limit obtained for an assumed excitation temperature of 168 K. \par

To account for the differences in methanol emission across our sample — for instance, due to a difference in emitting area — we adopted the \ce{CH3OH} abundance, defined as the ratio between the methanol and hydrogen column densities, as a common quantity for comparison. Since for most disks the methanol column density has been derived assuming an emitting area confined within the snowline, the corresponding hydrogen column density was also computed within this region, based on thermochemical models. For this purpose, we used published DALI \citep{DALI_1, DALI_2, DALI_3} models of these systems to estimate the average hydrogen column density ($N^{\rm snow}_{\rm h} = N^{\rm snow}_{\rm H} + 2N^{\rm snow}_{\rm H_2}$), present in both atomic and molecular form, of the gas inside the \ce{CH3OH} snowline ($r_{\rm snow}$). The procedure used to estimate $N^{\rm snow}_{\rm h}$ for each disk is described in App.~\ref{sec:Appendix_C}. For the HD~100453, CQ~Tau and DG~Tau~A disks, however, the hydrogen column density within the snowline could not be determined due to the absence of an available thermochemical model. Moreover, the TW~Hya disk was excluded from our analysis, as our assumption of thermally desorbed methanol does not apply to this source. In this case, the detected gas-phase methanol reservoir originates beyond the snowline, where thermal desorption is negligible, and is most likely released from dust grain surfaces via photodesorption \citep{Walsh_2016, Ilee+25}. \par

With the aid of Fig.~\ref{fig:meth_corr}, we search for trends between the $N_{\ce{CH3OH}}/N^{\rm snow}_{\rm h}$ ratio and the properties of the disk region where the thermally sublimated methanol reservoir is expected to reside. As proxies for these properties, we use the (sub-)millimeter continuum flux density (left panel) and the gas mass $M_{\rm snow}$ (right panel), both computed within the snowline. The method used to derive $M_{\rm snow}$ from the DALI thermochemical models is described in App.~\ref{sec:Appendix_C}, while App.~\ref{sec:Appendix_D} lists the ALMA Band~7 archival programs used to integrate the continuum flux inside $r_{\rm snow}$. All flux densities have been scaled to a common distance of 140~pc. \par

In both panels of Fig.~\ref{fig:meth_corr}, HL~Tau does not follow the trend found for the transition disks hosting Herbig stars, which exhibit a $N_{\ce{CH3OH}}/N^{\rm snow}_{\rm h}$ ratio at least three orders of magnitude higher than the upper limit derived for HL~Tau. The presence of a central cavity could explain why methanol and other COMs are detected in transition disks: stellar radiation directly irradiates and heats the inner edge of the cavity wall, releasing into the gas phase the material previously locked in the icy mantles of dust grains. Moreover, inside the cavity, the methanol emission is less severely obscured by dust. \par

The V883~Ori protoplanetary disk shows a $N_{\ce{CH3OH}}/N^{\rm snow}_{\rm h}$ ratio similar to that of other transition disks, but exhibits a much higher millimeter flux density and gas mass within the snowline. These values arise from the distinct thermal structure of the disk, which has been heated by a recent accretion outburst of the central FU~Orionis-type protostar, currently reaching a luminosity of $\sim$200~$L_{\odot}$ \citep{Pickering_1890, Furlan_2016}. As a consequence, the water snowline has shifted outward to $\sim$80~au from the star, effectively releasing the icy material from dust grains within this radius \citep{vantHoff_2018, Leemker_2021, Tobin_2023, Wang_2025}. \par 

Conversely, according to the $r_{\rm snow}$ estimated from the DALI thermochemical model, in the HL~Tau disk methanol should be thermally desorbed only in the inner disk within $\sim$4~au from the central star. Several multi-wavelength studies have shown that, at a frequency of $\sim$300~GHz, the dust in HL~Tau is still highly optically thick inside 60 au or inside 90 au, depending on the assumed grain porosity \citep{Gonzales+19, Guerra-Alvarado+24, Takahiro+25}. Optically thick dust can have a dramatic effect on the line emission \citep{Isella_2016, Weaver_2018, De_Simone+2020}, as it has also been shown for the V883~Ori protoplanetary disk, where the methanol and the water emission are not centrally peaked because the dust remains optically thick out to a radius of $\sim$40~au \citep{Cieza_2016, Houge_2024}. Therefore, optically thick dust can alter the line emission and thus the \ce{CH3OH} abundance derived from observations, possibly explaining the methanol non detection in HL~Tau. Nonetheless, a difference in the chemical composition of the disks in our sample cannot be excluded. Observations at longer wavelengths (e.g. cm) can further help to investigate the effect of dust attenuation on the COMs line emission at (sub-)millimeter wavelength \citep{De_Simone+2020}.

\subsection{Sources with methanol and sulfur monoxide emission}

Investigating the connection between the sulfur and the COMs chemistry in protoplanetary disks can open a window on the common condition that lead to the sublimation of these species from the ices. All the protoplanetary disks with methanol detections have also shown to host gas-phase sulfur monoxide reservoirs, indicating shocks or sufficiently high temperatures to allow \ce{SO} to desorb and/or to form through gas-phase chemistry reactions \citep{Van_Gelder_2021}. The latter is probably the case for HD~100546 \citep{Booth_2023_SO, Booth_2024}, HD~169142 \citep{Booth_2023, Law_2023}, HD~100453 \citep{Booth_2025b}, IRS~48 \citep{Booth_2024b, Temmink_2024}, CQ~Tau \citep{Booth_2025b, Zagaria_2025} and V883~Ori \citep{Ruiz_Rodriguez_2022}. Additionally, for the disks around HD~100546 and HD~169142, SO could also potentially trace the presence of an embedded planet. In all these sources, the \ce{CH3OH} and the \ce{SO} emission originate from a similar region of the disk enclosed within the water snowline. In TW~Hya, instead, the \ce{SO} emission is asymmetric and localized in the southeast region of the disk, most likely due to an outflow driven by an embedded proto-planet in one of the gaps \citep{Yoshida_2024}. Several sulfur monoxide lines have also been detected in the HL Tau system, tracing both the extended streamer seen in \ce{HCO+} and the accretion shock where the streamer impacts the disk \citep{Garufi_2022, Leemker_2025_A&A}. Moreover, as probed by the higher-excitation \ce{SO} transitions, sulfur monoxide emission also originates from the inner disk of HL~Tau, likely from \ce{SO} molecules spiraling toward the star after being released from dust grains in the accretion shock. Similarly, volatile SO and \ce{SO2} are released into the gas phase on the north-west side of the DG~Tau~A disk, at the intersection with the streamer \citep{Garufi_2022}.  \par

In order to compare the methanol and the sulfur monoxide emission in the disks where the two molecules present a similar spatial distribution, we followed the methodology described by \cite{Zagaria_2025}, rescaling both the integrated \ce{CH3OH} and \ce{SO} line fluxes to the same transition and to the same distance of 140 pc according to eq.~\ref{eq:scaling}. The chosen reference lines are the $J_{N}=6_{5}-5_{4}$ \ce{SO} transition (as in \citealt{Zagaria_2025}) and the $6_{(1,5)}-6_{(0,6)}$ A \ce{CH3OH} line, since it returns the most stringent upper limit on the methanol column density in HL Tau. We assumed an LTE temperature of 50 K for SO, while we used as $T_{\rm ex}$ for methanol the values present in Tab.~\ref{tab:meth_in_disk} for each disk. For HL Tau, we considered the disk integrated \ce{SO} line flux from the high upper energy level $J_{N}=15_{15}-14_{14}$ transition presented in \cite{Garufi_2022}, since the emission is mostly coming from the central part of the disk where \ce{CH3OH} should be thermally desorbed. The fluxes of the chosen methanol and sulfur monoxide transitions, as well their rescaled values, are reported for each disk in Table \ref{tab:meth_and_so}. Due to the different morphology of the \ce{CH3OH} and \ce{SO} emission in the disk, TW~Hya has not been considered in our analysis. The same consideration also applies to DG~Tau~A, where the SO emission originates from the north-west region of the disk, while the (non-detected) thermally desorbed methanol reservoir is thought to reside in the inner disk. \par

\begin{figure} [htb!]
\centering
\includegraphics[width = 1 \linewidth]{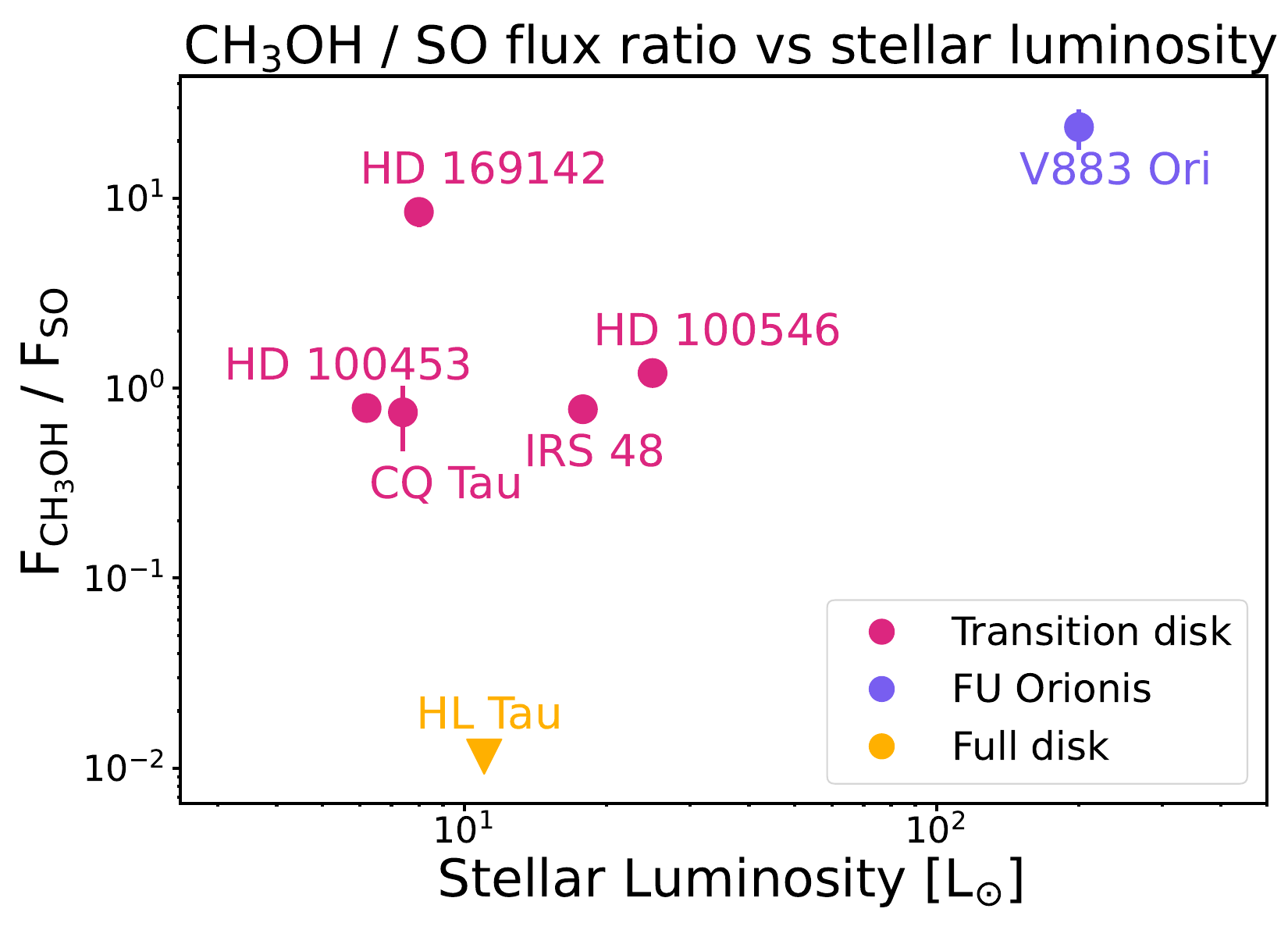}
\caption{\small{Methanol-to-sulfur monoxide line flux ratio as a function of the stellar luminosity. In addition to HL Tau,} we took into consideration all the protoplanetary disks in which \ce{SO} and \ce{CH3OH} are both detected and emitting from a similar region of the disk enclosed within the water snowline.}
\label{fig:meth_SO}
\end{figure}

Figure~\ref{fig:meth_SO} shows how the methanol-to-sulfur monoxide flux ratio varies as a function of stellar luminosity for the disks in our sample. No correlation is observed. Interestingly, once again, the ratio for HL Tau is at least one order of magnitude lower than that in other protoplanetary disks, suggesting the possibility that the inner disk where both chemical species are thermally desorbed from the dust grains is depleted in gas-phase methanol. Another possible explanation is that the measured ratio for HL~Tau is actually representative of the chemical composition of the region where the streamer impacts the disk. Indeed, even if the emission from the high-excitation \ce{SO} transitions originates from the inner disk, the bulk of this volatile sulfur-bearing molecule reservoir likely comes from \ce{SO} liberated by the accretion shock and then rapidly redistributed near the protostar. Since the estimated drifting timescale is shorter than the chemical reprocessing timescale required to alter the \ce{SO} and \ce{SO2} abundances set by the accretion shock \citep{Garufi_2022}, the upper limit on the methanol-to-sulfur monoxide flux ratio in HL~Tau would trace the chemical composition of the region where the streamer impacts the disk, under the assumption that also methanol is released into the gas-phase from the shock and transported inward as \ce{SO}.

\section{Summary}\label{sec:conclusion}

In this work we have presented the analysis of ALMA archival data of the HL Tau protoplanetary disk covering bright predicted methanol transitions. Our main findings are:

\begin{enumerate}
    \item We did not detect methanol in any of the datasets. Under the assumption that gas-phase methanol is thermalized and emitting within the water snowline, we derived the most stringent upper limit up to date on the \ce{CH3OH} column density in this disk (< $7.2 \times 10^{14} \ \mathrm{cm^{-2}}$ at 100 K and < $1.8 \times 10^{15} \ \mathrm{cm^{-2}}$ at 200 K). Furthermore, the low \ce{CH3OH} abundance compared to transition disks and V883~Ori suggests that the bulk of gas-phase methanol is hidden by optically thick dust in the HL Tau disk.
    \item We computed the upper limit on the methanol-to-water column density ratio, finding a value of < $0.55 \times 10^{-3}$ at 100 K and < $1.4 \times 10^{-3}$ at 200 K, which are one order of magnitude lower than the one for other YSOs. 
    \item In the region where the streamer impacts the disk, the methanol-to-sulfur monoxide flux ratio for HL Tau is at least two orders of magnitude lower than the one in other disks with \ce{CH3OH} detections.
    \item We showed that, when searching for weak lines on top of a bright continuum source, special care must be taken to optimize the signal-to-noise ratio per channel for the bandpass calibrator, not to impart any noise into the scientific target.
\end{enumerate}



\begin{acknowledgements}
We are grateful to the anonymous referee for the constructive comments and suggestions. We thank Luna Rampinelli for the useful discussions and Luke Keyte for sharing the HD~169142 disk model. \par
This paper makes use of the following ALMA data: 2012.1.00799.S, 2013.1.00100.S, 2015.1.00806.S, 2016.1.00629.S, 2016.1.00728.S, 2017.1.01178.S, 2019.1.00393.S and 2022.1.00905.S. ALMA is a partnership of ESO (representing its member states), NSF (USA) and NINS (Japan), together with NRC (Canada), NSTC and ASIAA (Taiwan), and KASI (Republic of Korea), in cooperation with the Republic of Chile. The Joint ALMA Observatory is operated by ESO, AUI/NRAO and NAOJ. \par
A.S. and L.T work was partly supported by the Italian Ministero dell'Istruzione, Universit\`a e Ricerca through the grant Progetti Premiali 2012 – iALMA (CUP C$52$I$13000140001$).
This project has received funding from the European Union's Horizon 2020 research and innovation programme through the  European Research Council (ERC)  ERC Synergy Grant {\em ECOGAL} (grant 855130). Views and opinions expressed are however those of the author(s) only and do not necessarily reflect those of the European Union or the European Research Council Executive Agency. Neither the European Union nor the granting authority can be held responsible for them. \par
M.L. and S.F. are funded by the European Union (ERC, UNVEIL, 101076613). Views and opinions expressed are however those of the author(s) only and do not necessarily reflect those of the European Union or the European Research Council. Neither the European Union nor the granting authority can be held responsible for them. S.F. acknowledges financial contribution from PRIN-MUR 2022YP5ACE.  \par

\end{acknowledgements}

%
%

\bibliography{biblio} 


\appendix 

\section{Calibration of the observations}\label{sec:Appendix_A} 

The dataset covering the selected \ce{CH3OH} lines belong to Band 7 ALMA observations from three programs: 2017.1.01178.S (PI: Humphreys, E.), 2022.1.00905.S (PI: Facchini, S.) and 2019.1.00393.S (PI: Zhang, K.). Within the program 2017.1.01178.S, HL Tau has been observed in four execution blocks, two of which being in Band 7. The first Band 7 execution block was observed on November 24, 2017, with an integration time of 34 minutes and with an array configuration of 49 antennas, whose baselines ranged from 92 m to 8.5 km. J0431+1731 was used for phase calibration, and J0538-4405 for bandpass and flux calibration. The second execution block was observed for 31 minutes on August 12, 2019. In this case, 48 antennas with a maximum separation of 3.6 km were employed. The phase calibration and the flux/bandpass calibration relied, respectively, on the sources J0440+1437 and J0519-4546. The chosen spectral set-up, consisting in four spectral windows (spws) in FDM mode, was similar for both observations. Three of the spws have 1920 channels with a width of 976 kHz ($\sim$ 0.9 km s$^{-1}$), for a total 1.875 GHz bandwidth, while the other one was set to an higher spectral resolution of 244 kHz ($\sim$ 0.2 km s$^{-1}$). \par
Within the program 2022.1.00905.S, HL Tau has been targeted by 41/45 antennas during two execution blocks observed in October 12 (EB0) and 15 (EB1), 2022, for a total integration time of 100 minutes. The maximum baseline was of 500 m, providing a more moderate resolution of $\sim 0.8''$ when compared to those achieved by the other two programs. The spectral set-up consisted of two low resolution spws with a channel width of 976 kHz ($\sim$ 0.9 km s$^{-1}$), and in two high resolution spws with a channel width of 244 kHz ($\sim$ 0.2 km s$^{-1}$), for a bandwidth of, respectively, 1.875 and 0.469 GHz. J0423-0120 was used as flux and bandpass calibrator, while J0431+1731 was used for phase referencing. More specific information on the self-calibration performed combining together the B7 datasets from program 2017.1.01178.S and 2022.1.00905.S are included in \cite{Facchini_2024}. \par
Lastly, the two execution blocks from program 2019.1.00393.S took place on January 3 and on May 5, 2022, for a total integration time of 43 minutes. The array configuration was comprised of 38/39 antennas, with baselines ranging from 15 m to 782 m. Out of the six spws, three were configured with a high spectral resolution of 122 kHz ($\sim$ 0.1 km s$^{-1}$), while the channel width for the remaining ones was of 244 and 976 kHz. J0510+1800 was used for flux and bandpass calibration, while J0431+1731 was used for phase calibration. 

\section{Calibration of program 2022.1.00905.S observations}\label{sec:Appendix_B}
 
In searching for weak lines in targets with strong continuum emission -such as is the case for the HL Tau protoplanetary disk-, ideally one would need to achieve a bandpass signal-to-noise-ratio (SNR) at least five times higher than the expected target SNR per channel. Otherwise the SNR of the bandpass will impose a spectral dynamic range limit, meaning that weak lines will not be detectable given the bandpass solution SNR with respect to the target source strength. \par

This issue is evident in the observations from program 2022.1.00905.S. To illustrate it, we first \textit{tclean}ed HL Tau with a 2 km s$^{-1}$ channel spacing and with a \verb+robust+ parameter of 2. The \verb+auto-multithresh+ algorithm \citep{Kepley_2020} was used to clean down to a $3\sigma$ level. The image cube indicates a peak flux of $\sim$ 0.8 Jy~beam$^{-1}$ per channel with a noise of $\sim$ 0.7 mJy~beam$^{-1}$ per channel, i.e. a spectral SNR in the image of $\sim$ 1100. Then we also \textit{tclean}ed the bandpass calibrator using the same parameters, achieving a spectral SNR of $\sim$ 2000 (given the peak flux of $\sim$ 3 Jy~beam$^{-1}$ per channel and a rms of $\sim$ 1.5 mJy~beam$^{-1}$ per channel), which is just two times higher than that of the science target when solving for the bandpass at the same spectral resolution. Likewise, the antenna based solutions are of the same ratio. This would effectively hide any weak line in the noise. \par

\begin{figure*}[htb!]
\centering
\includegraphics[width = 0.9 \linewidth]{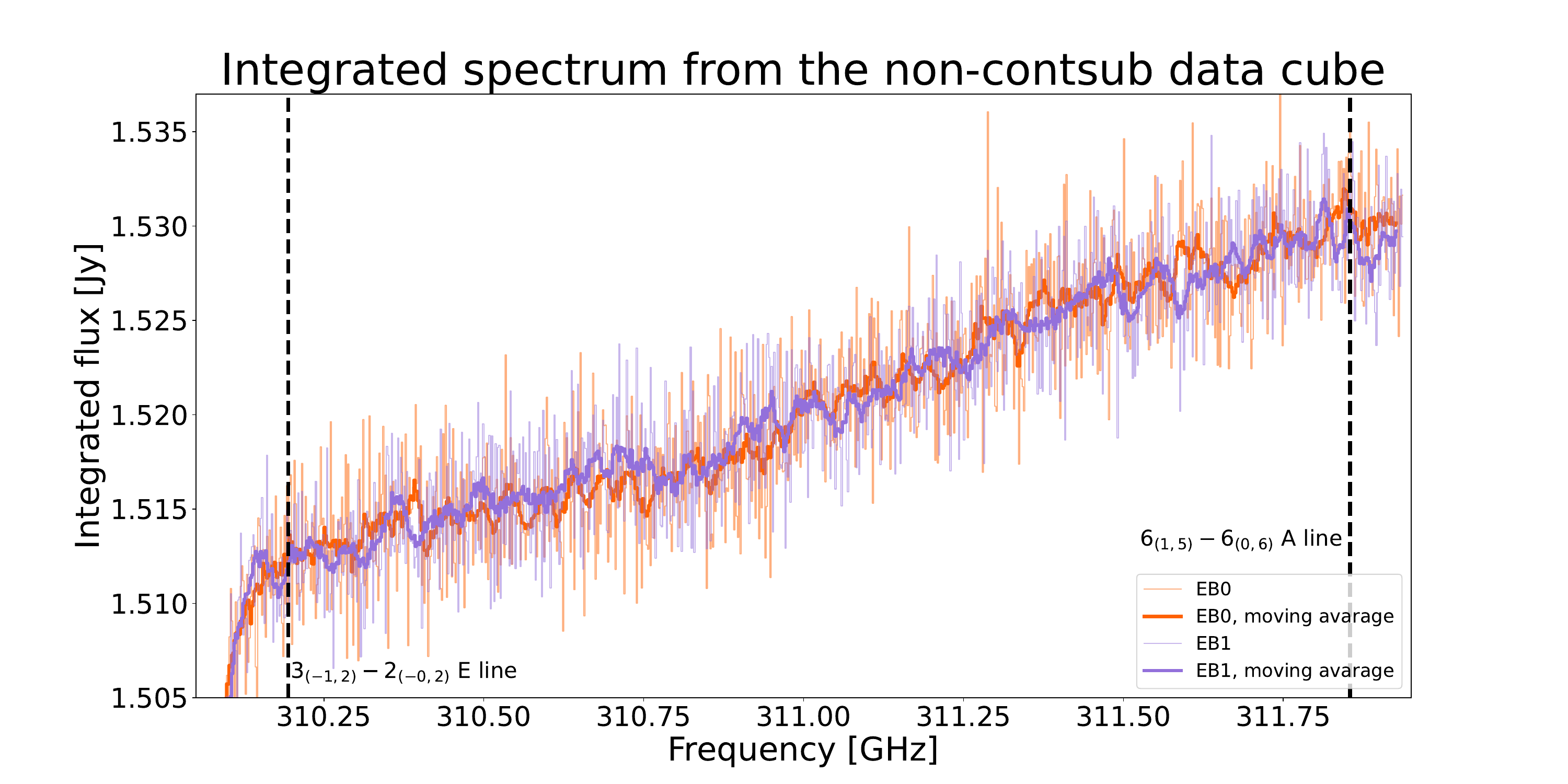}
\caption{\small{Integrated spectrum from a 1$''$ circular region centered on HL Tau and extracted from the non-continuum subtracted data cube. The orange and purple line differentiate the spectra extracted in two different execution blocks of the same spectral window. The bold lines were computed performing a moving average over ten consecutive channels to highlight the variations on a scale of $\sim$ 0.003 Jy. The black dotted lines indicate the region of the spectrum where the two covered and non-detected methanol $3_{(-1,2)}-2_{(-0,2)}$ E and $6_{(1,5)}-6_{(0,6)}$ A line fall.}}
\label{fig:spw8_12}
\end{figure*}

Fig.~\ref{fig:spw8_12} illustrates how this limited dynamic range affects the broad band observations from program 2022.1.00905.S. It shows the integrated spectrum from the image cube extracted over a 1$''$ circular area centered on HL Tau for the same spectral window (spw) observed in the two execution blocks EB0 and EB1. The native pipeline reduction is set to achieve at least an antenna based SNR of 50 per channel, and in the case of the wide band 1.875 GHz spw we use, there is no channel averaging as each channel has an estimated 170 SNR. The moving average performed on the integrated spectra from the image cube highlights a flux variation of $\sim$ 0.003 Jy, effectively hindering a robust detection of any weak line. \par

To boost the dynamical range, we performed a new bandpass calibration binning together 38 channels per solution, improving six times the previous reached bandpass signal-to-noise ratio per channel, which should be sufficient for the bandpass noise not to limit the target. However, as the attempted calibration with more channel binning does not remove the intrinsic bandpass features on the source spectra, we decided to resort for our analysis to the original standard ALMA pipeline calibration, motivated also by the little to no improvement in the HL Tau image cube noise. The only sure way to get better SNR on the bandpass is to try select brighter sources or observe it for longer. Of course this come with the acknowledgment that the observations will be a less efficient, but could be necessary for such difficult weak line searches. Nonetheless, when carefully inspecting the spectra from these 2022.1.00905.S. program datasets, we took into consideration the limited dynamic range, in order to avoid mistaking a weak line for an induced instrumental ripple.

\section{Computing $N_{\rm h}$ and $M_{\rm snow}$ from DALI thermochemical model}\label{sec:Appendix_C}

\begin{table*}[htb!]
    \centering
    \caption{This tables summarizes the relevant disk and stellar parameters of all disks with \ce{CH3OH} detections.}
    \addtolength{\leftskip} {-2cm} 
    \addtolength{\rightskip}{-2cm}
    \begin{tabular}{p{0.19\columnwidth}p{0.042\columnwidth}p{0.042\columnwidth}p{0.042\columnwidth}cp{0.052\columnwidth}p{0.042\columnwidth}p{0.08\columnwidth}p{0.042\columnwidth}p{0.042\columnwidth}cp{0.042\columnwidth}cc}
        \hline
        Name & d & $M_{*}$ & $L_{*}$ & $N_{\ce{CH3OH}}$ & $T_{\rm ex}^{(a)}$ & $r_{\rm snow}$ & $\Sigma_{0}$ & $R_{\rm c}$ & $\gamma$ & $\delta_{\rm gas}$ & $R_{\rm gas,cav}$ & $\mathrm{M_{\rm snow}}$ & Refs.$^{(b)}$ \\
         & [pc] & [$M_{\odot}$] & [$L_{\odot}$] & [cm$^{-2}$] & [K] & [au] & [g/cm$^{2}$] & [au] &  &  & [au] & [$M_{\odot}$] & \\
        \hline
        TW Hya & 60 & 0.8 & 0.3 & $1.8^{+1.3}_{-0.5}\times10^{12}$& 36 & 2.0 & 30 & 35 & 1 & $10^{-2}$ & 4 & 1.3$\times10^{-5}$  & (1,2) \\
        HD 100546 & 110 & 2.1 & 25 & (1.4$\pm$0.4)$\times10^{14}$ & 152 & 15 & 250 & 30 & 1 & $10^{-5}$,$10^{-2}$ & 15 & 1.7$\times10^{-6}$ & (3,4) \\
        HD 169142 & 114 & 1.7 & 8 & (3.3$\pm$0.3)$\times10^{14}$ & 100* & 21 & 7 & 100 & 1 & $10^{-10}$,$10^{-3}$ & 13 & 3.9$\times10^{-6}$ & (5,6) \\
        HD 100453 & 104 & 1.5 & 6 & $1.6^{+0.2}_{-0.1}\times10^{14}$ & 228 & / & / & / & / & / & / & / & (7)\\
        IRS 48 & 134 & 2.0 & 18 & (4.9$\pm$0.2)$\times10^{14}$ & 103 & 60 & 0.3 & 60 & 1 & $10^{-3}$ & 25 & 1.7 $\times$ $10^{-4}$ & (8,9) \\
        CQ Tau & 149 & 1.5 & 7 & -$^{(c)}$ & / & / & / & / & / & / & / & /  & / \\
        V883 Ori & 388 & 1.3 & 200 & $4.8^{+0.5}_{-0.5}\times10^{17}$ & 92 & 78 & 35 & 75 & 1 & \multicolumn{2}{c}{/} & 8.7  $\times$ $10^{-2}$ & (10,11) \\
        HL Tau & 140 & 2.1 & 11 & < 1.4$\times10^{15}$ & 168* & 4 & 314 & 30 & 1 & \multicolumn{2}{c}{/} & 2.3 $\times$ $10^{-2}$ & (12,13)\\
        \hline 
    \end{tabular}
    \tablefoot{Distances of the source, stellar masses and luminosities are taken from the references listed in \cite{Francis_Nienke_2020}, with the exception of CQ~Tau \citep{GAIA, Vioque_2018}, V883~Ori \citep{Furlan_2016}, HL~Tau \citep{Liu_2017, Yen_2019}. $^{(a)}$ An asterisk beside $T_{\rm ex}$ indicates that the temperature was assumed, rather then measured via a rotational diagram. $^{(b)}$
    The first column lists the references from which the methanol column densities and the methanol excitation temperatures are taken from. The second column contains the publications where the DALI models parameters are described.  Additionally, other thermochemical models are presented in: HD~100546 \citep{Keyte+2023} and IRS~48 \citep{Bruderer_2014, VanderMarel+16, VanDerMarel_2021}. No thermochemical model was available for the disk HD~100453. $^{(c)}$ Due to the detection of just one methanol line in the CQ~Tau disk, \cite{Booth_2025b} report only the value of the integrated \ce{CH3OH} line flux.}
    \tablebib{(1)~\cite{Ilee+25};
    (2)~\cite{Kama_2016};
    (3)~\cite{Evans_2025};
    (4)~\cite{Leemker_2024};
    (5)~\cite{Booth_2023};
    (6)~\cite{Keyte+24};
    (7)~\cite{Booth_2025a};
    (8)~\cite{VanDerMarel_2021};
    (9)~\cite{Leemker_2023};
    (10)~\cite{Tobin_2023};
    (11)~\cite{Leemker_2021};
    (12)~This work;
    (13)~\cite{Leemker_2025_A&A};
    }
    \label{tab:meth_in_disk}
\end{table*}

In this section, we will explain the procedure that we followed to compute the average hydrogen column density $N^{\rm snow}_{\rm h}$, present in both atomic and molecular form, and the gas mass $M_{\rm snow}$ inside the \ce{CH3OH} snowline $r_{\rm snow}$. Firstly, we estimated  $r_{\rm snow}$ as the midplane location where the dust temperature reaches the methanol desorption temperature $\sim 120$ K. Following the prescription from \cite{Lynden-Bell_1974} and \cite{Hartmann_1998}, the radial gas surface density $\Sigma(R)$ profile used for the disk models is given by the product between a power law and a negative, outer disk dominating exponential:

\begin{equation}\label{eq:gas_surface}
    \Sigma(R) = \Sigma_{0} \times \left( \frac{R}{R_{\rm c}}\right)^{-\gamma} \times \exp \left[- \left( \frac{R}{R_{\rm c}}\right)^{2 -\gamma} \right].
\end{equation}

In the expression above, $\Sigma_{0}$ is the surface density at the characteristic radius $R_{c}$, while $\gamma$ is the power law index. To account for the central cavity of the transitional disks in our sample, we multiplied the surface density in eq. \ref{eq:gas_surface} by the $\delta_{\rm gas}$ parameter, which describes the relative drop in gas density inside the central gap, up to the outer radius of the gas cavity $R_{\rm gas,cav}$. We have used a secondary $\delta_{\rm gas}$ of $10^{-2}$ and $10^{-3}$ when integrating $\Sigma$ between $R_{\rm gas,cav}$ and the outer radius of the dust cavity $R_{\rm dust,cav}$ for, respectively, the HD~100546 and HD~169142 disks following the models presented in \cite{Leemker_2024} and \cite{Keyte+24}. For both of them, $r_{\rm snow}$ is locate between $R_{\rm gas,cav}$ and $R_{\rm dust,cav}$.
A summary of the relevant model parameters and references to the original works are presented in Table~\ref{tab:meth_in_disk}, together with the computed values of $r_{\rm snow}$ and $M_{\rm snow}$. \par

We estimate $M_{\rm snow}$ by integrating the surface density self-similar solution of a viscously evolving disk up to the snowline:

\begin{equation}\label{eq:M_snow}
    {M_{\rm snow}} =  \int^{2\pi}_{0} \int^{r_{\rm snow}}_{0} R\Sigma(R)dRd\phi  = 2\pi m_{\rm p} \int^{r_{\rm snow}}_{0} RN(R)dR,
\end{equation}

where $m_{\rm p}$ is the mean mass of one particle in the DALI model ($1.4 \times1.667 \times10^{-24}$g), while $N(R)$ is the gas column density. Similarly, the average hydrogen column density within the snowline $N^{\rm snow}_{\rm h}$ was computed as:

\begin{equation}\label{eq:M_snow}
    {N^{\rm snow}_{\rm h}} = \frac{2\pi \int^{r_{\rm snow}}_{0} RN_{\rm h}(R)dR}{\pi r^2_{\rm snow}},
\end{equation}

where $N_{\rm h}(R)$ is the hydrogen column density as a function of the radius. \par 

\setcounter{table}{0}
\renewcommand{\thetable}{E.\arabic{table}}

\begin{table*}[htb!]
    \centering
    \caption{\ce{CH3OH} and \ce{SO} line fluxes for the disks with methanol detections}
    \addtolength{\leftskip} {-2cm} 
    \addtolength{\rightskip}{-2cm}
    \begin{tabular}{p{0.19\columnwidth}p{0.2\columnwidth}p{0.14\columnwidth}p{0.21\columnwidth}p{0.22\columnwidth}p{0.18\columnwidth}p{0.14\columnwidth}p{0.22\columnwidth}p{0.22\columnwidth}p{0.07\columnwidth}}
        \hline
        Name & \ce{CH3OH} & $\nu_{\ce{CH3OH}}$ & Obs. flux & Resc. flux$^{(a)}$ & \ce{SO} & $\nu_{\ce{SO}}$ & Obs. flux & Resc. flux$^{(a)}$ & Refs.$^{(b)}$\\
        & Transition & [GHz] & [mJy km $\rm s^{-1}$] & [mJy km $\rm s^{-1}$] & Transition & [GHz] & [mJy km $\rm s^{-1}$] & [mJy km $\rm s^{-1}$] &\\
        \hline
        HD 100546 & $7_{(0,7)}$-$6_{(0,6)}$ A & 338.4087 & 137 $\pm$ 19 & 149 $\pm$ 21 & $7_{8}-6_{7}$ & 340.7142 & 343 $\pm$ 26 & 124 $\pm$ 9 & (1,2) \\
        HD 169142 & $4_{(2,2)}$-$3_{(1,2)}$ E & 218.4401 & 65 $\pm$ 7 & 379 $\pm$ 41 & $8_{8}-7_{7}$ & 344.3106 & 120 $\pm$ 16 & 45 $\pm$ 6 & (3,4)\\
        HD 100453 & $6_{(1,5)}$-$6_{(0,6)}$ A & 311.8526 & 141 $\pm$ 8 & 78 $\pm$ 5 & $7_{7}-6_{6}$ & 301.2861 & 258 $\pm$ 5 & 99 $\pm$ 2 & (5,6)\\
        IRS 48 & $7_{(1,7)}$-$6_{(1,6)}$ E & 338.3446 & 67 $\pm$ 4 & 441 $\pm$ 21 & $7_{8}-6_{7}$ & 340.7142 &  1063 $\pm$ 23 & 569 $\pm$ 12 & (7,8)\\
        CQ Tau & $6_{(1,5)}$-$6_{(0,6)}$ A & 311.8526 & 54 $\pm$ 17 & 61 $\pm$ 19 & $6_{5}-5_{4}$ & 219.9494 & 72 $\pm$ 15 & 82 $\pm$ 17 & (9,10)\\
        V883 Ori$^{(c)}$ & $5_{(4,2)}$-$6_{(3,3)}$ A& 364.2027 & 800 $\pm$ 180 & (2.0$\pm$0.4)$\times 10^{5}$ & $6_{5}-5_{4}$ & 219.9494 & 1100 $\pm$ 100 & 8449 $\pm$ 768 & (11,12)\\
        HL Tau & $6_{(1,5)}$-$6_{(0,6)}$ A & 311.8526 & < 10 & < 10 & $15_{15}-14_{14}$ & 645.2549 & 591 $\pm$ 90 & 855 $\pm$ 130 & (13,14)\\
        \hline   
    \end{tabular}
    \tablefoot{All the observed line fluxes are disk-integrated, with the exception for those taken from the following references: \cite{Ruiz_Rodriguez_2022,vantHoff_2018, Zagaria_2025}. $^{(a)}$The observed \ce{CH3OH} and \ce{SO} line fluxes have been rescaled to a common distance of 140 pc and to the same transition (the $6_{(1,5)}-6_{(0,6)}$ A one for \ce{CH3OH} and the $J_{N}=6_{5}-5_{4}$ transition for \ce{SO}). $^{(b)}$In the column, the first entry points to the reference where the \ce{CH3OH} line flux values are presented, while the second entry does the same for \ce{SO}. $^{(c)}$The high rescaled \ce{CH3OH} and \ce{SO} fluxes inside the V883 Ori disk suggest that the emission could be optically thick. To be consistent with the other sources, we still rescale the fluxes using eq. \ref{eq:scaling} under the assumption of optically thin emission.}
    \tablebib{{(1)~\cite{Evans_2025}};
    (2)~\cite{Booth_2024};
    (3)~\cite{Booth_2023};
    (4)~\cite{Law_2023};
    (5)~\cite{Booth_2025a};
    (6)~\cite{Booth_2025b};
    (7)~\cite{Temmink_2024};
    (8)~\cite{Booth_2024b};
    (9)~\cite{Booth_2025b};
    (10)~\cite{Zagaria_2025};
    (11)~\cite{vantHoff_2018};
    (12)~\cite{Ruiz_Rodriguez_2022};
    (13)~This work;
    (14)~\cite{Garufi_2022};
    }
    \label{tab:meth_and_so}
\end{table*}
\section{ALMA archival observation for the disk continuum flux}\label{sec:Appendix_D}

In order to compute the continuum flux density at $\sim$300 GHz within the methanol snowline, we took available continuum product data from the ALMA archive, choosing for each disk B7 observations with good angular resolution (< 0.2$''$, in order to spatially resolve the snowline) and continuum sensitivity (< 0.06 mJy). The chosen programs from each disk are: HD~100546 (2015.1.00806.S; PI: Pineda, J.; \citealt{HD_100546_cont}), HD~169142 (2012.1.00799.S; PI: Honda, M.; \citealt{HD_169142_cont}), IRS~48 (2013.1.00100.S; PI: van der Marel, N.; \citealt{VanderMarel+16}), V883~Ori (2016.1.00728.S; PI: Cieza, L.; \citealt{Lee_2019}). For HL Tau we imaged the continuum from the datasets from program 2017.1.01178.S. \par

\section{\ce{CH3OH} and \ce{SO} line fluxes}\label{sec:Appendix_E}

Table \ref{tab:meth_and_so} presents quantum numbers, frequencies, measured and rescaled fluxes for the \ce{CH3OH} and \ce{SO} transitions that have been considered to compute the \ce{CH3OH}/\ce{SO} line flux ratio shown in Fig.~\ref{fig:meth_SO}. Under the assumption of optically thin emission, we rescaled the observed line fluxes $F_{\rm obs}$ to a common distance of 140 pc and to the the same transition, according to the following equation \citep{Rot_Diagram}:

\begin{equation}\label{eq:scaling}
    F_{\rm ref} = F_{\rm obs} \cdot \left( \frac{\rm distance}{140 \ \rm pc} \right)^2 \cdot \frac{A_{\rm ul}^{\rm ref}g_{u}^{\rm ref}}{A_{\rm ul}^{\rm obs}g_{u}^{\rm obs}} \cdot \exp  \left(-\frac{(E_{\rm u}^{\rm ref}-E_{\rm u}^{\rm obs})}{kT_{\rm ex}} \right)
\end{equation}

\section{\ce{CH3OH} integrated spectra}\label{sec:Appendix_F}

\begin{figure*} [htb!]
\centering
\includegraphics[width = 0.8 \linewidth]{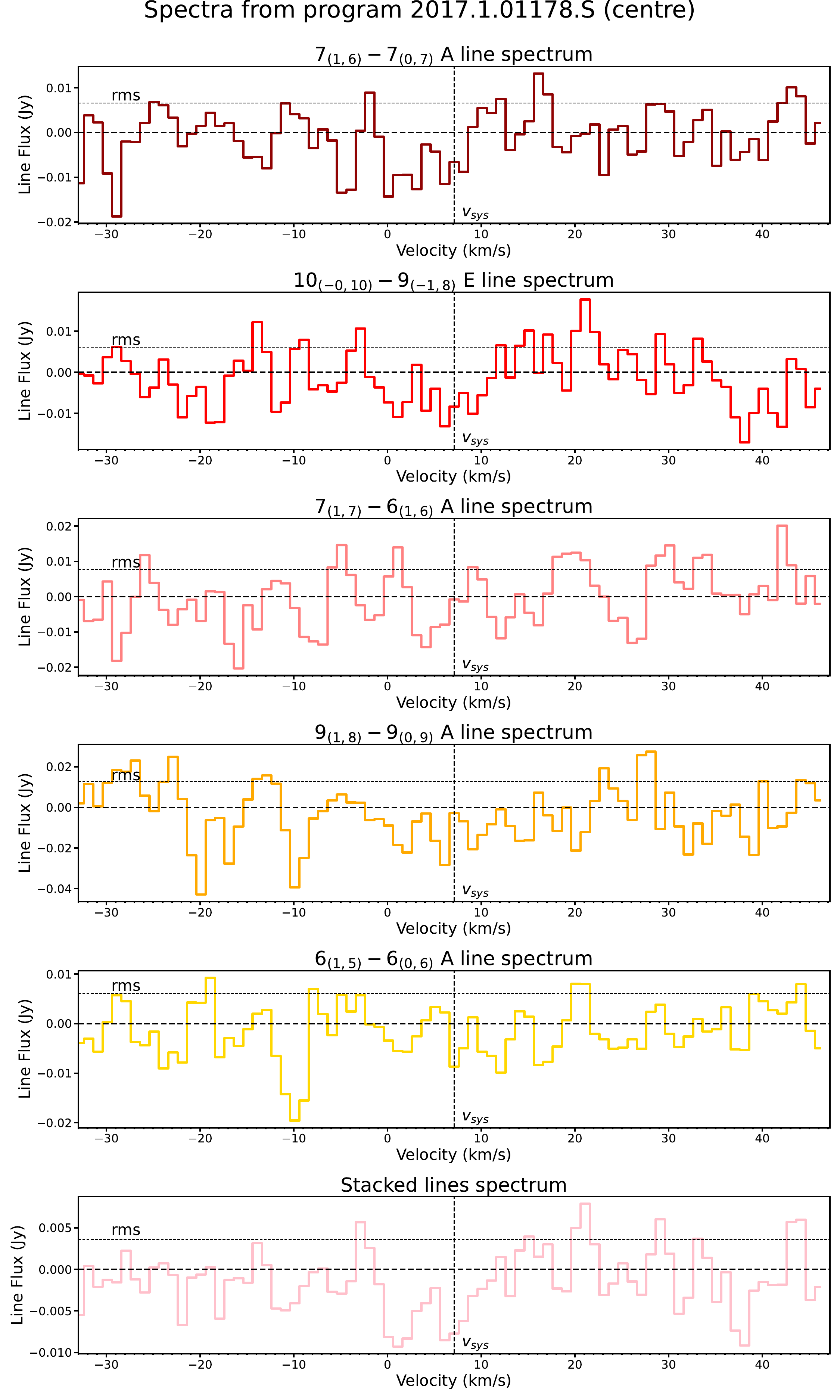}
\caption{\small{Spectra extracted for all the \ce{CH3OH} lines covered inside the program 2017.1.01178.S}}
\label{fig:H_spectra}
\end{figure*}

\begin{figure*} [htb!]
\centering
\includegraphics[width = 1. \linewidth]{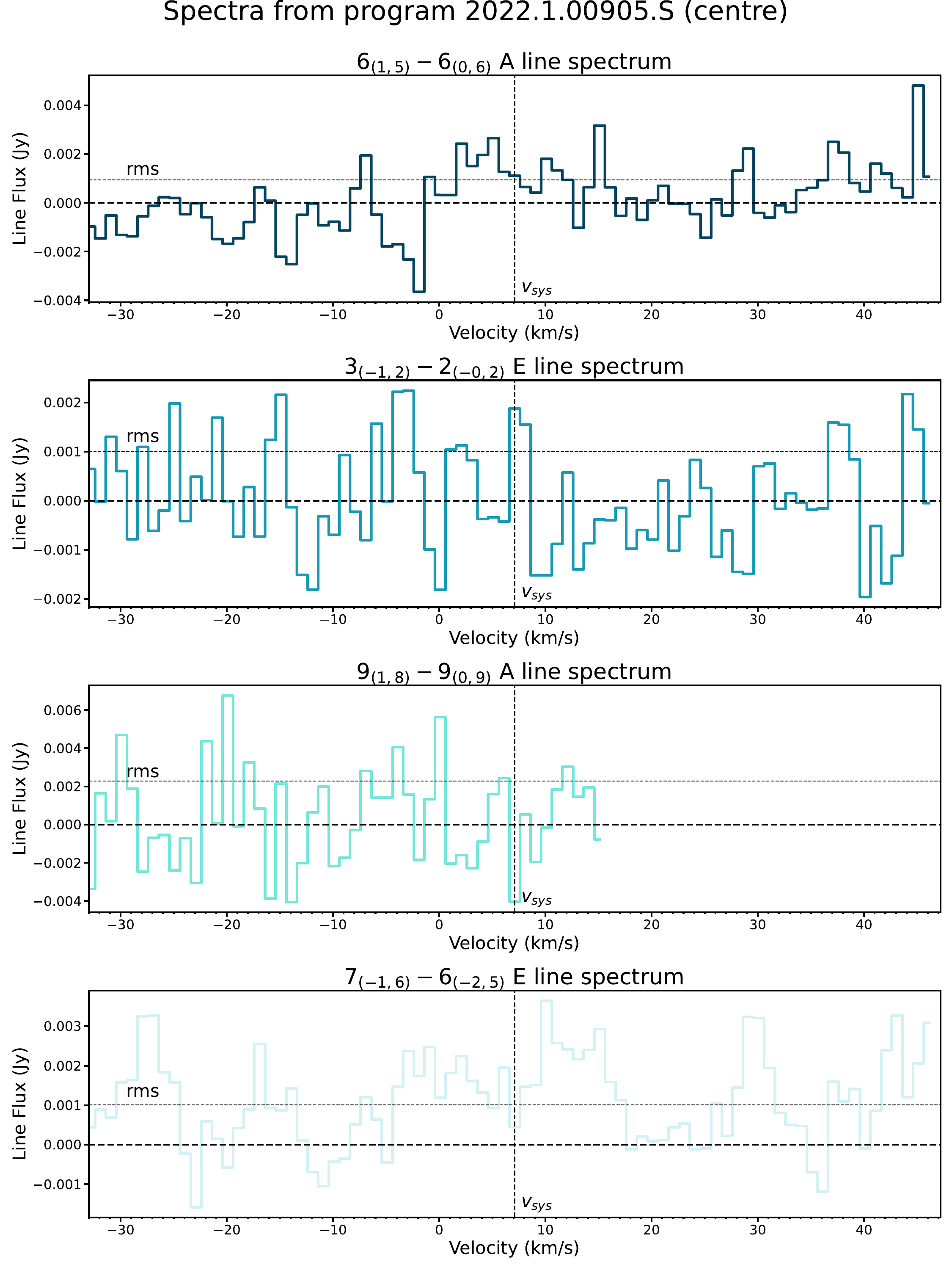}
\caption{\small{Spectra extracted for all the \ce{CH3OH} lines covered inside the program 2022.1.00905.S.}}
\label{fig:S_spectra}
\end{figure*}

\begin{figure*} [htb!]
\centering
\includegraphics[width = 1. \linewidth]{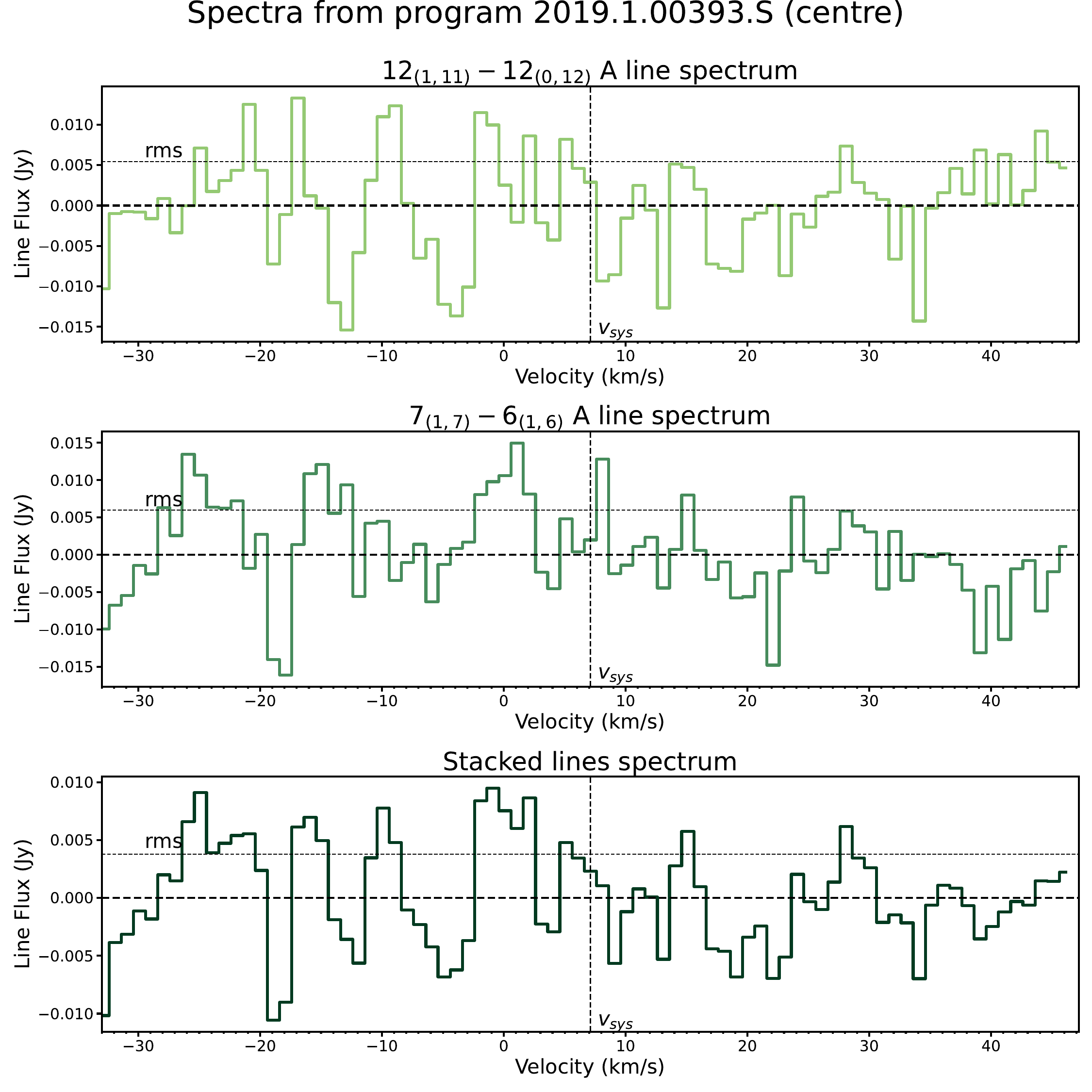}
\caption{\small{Spectra extracted for all the \ce{CH3OH} lines covered inside the program 2019.1.00393.S.}}
\label{fig:Z_spectra}
\end{figure*}

Figures \ref{fig:H_spectra}, \ref{fig:S_spectra}, \ref{fig:Z_spectra} display, respectively, the spectra from the lines covered in the programs 2017.1.01178.S, 2022.1.00905.S, and 2019.1.00393.S. These are all integrated spectra extracted from a circular area centered on HL Tau with a radius of 0.7$''$.

\end{document}